\documentclass[usenatbib,usegraphicx]{mn2e} 

\newcommand{\chandra}{\textit{Chandra}}
\newcommand{\rosat}{\textit{ROSAT}}
\newcommand{\iue}{\textit{IUE}}
\newcommand{\xmm}{\textit{XMM-Newton}}
\newcommand{\zpup}{$\zeta$~Pup}
\newcommand{\zori}{$\zeta$~Ori}

\newcommand{\tsco}{$\tau$~Sco}
\newcommand{\bcru}{$\beta$~Cru}
\newcommand{\bcep}{$\beta$~Cephei}

\newcommand{\Rstar}{\ensuremath{\rm R_{\ast}}}
\newcommand{\Rmin}{\ensuremath{R_{\rm min}}}
\newcommand{\Rfir}{\ensuremath{R_{\rm fir}}}
\newcommand{\Rsun}{\mathrm {R_{\sun}}}
\newcommand{\Lsun}{\mathrm {L_{\sun}}}
\newcommand{\Msun}{\mathrm {M_{\sun}}}
\newcommand{\Msunyr}{\ensuremath{\mathrm {M_{\sun}~{\mathrm yr^{-1}}}}}
\newcommand{\hetgs}{HETGS}

\newcommand{\kms}{km s$^{-1}$}
\newcommand{\vhwhm}{\ensuremath{v_{\mathrm {hwhm}}}}
\newcommand{\ftoi}{\ensuremath{\mathit f/i}}

\newcommand{\EM}{\ensuremath{\mathcal EM}}
\newcommand{\vinf}{\ensuremath{v_{\infty}}}
\newcommand{\vhundred}{\ensuremath{v_{\rm 100}}}
\newcommand{\Teff}{\ensuremath{T_{\rm eff}}}
\newcommand{\lx}{\ensuremath{L_{\rm x}}}

\newcommand{\mdot}{\ensuremath{\dot{M}}}
\newcommand{\Mdot}{\ensuremath{\dot{M}}}
\newcommand{\Mdotnine}{\ensuremath{\dot{M}_{-9}}}
\newcommand{\apj}{ApJ}
\newcommand{\apjs}{ApJ}
\newcommand{\apjl}{ApJ}
\newcommand{\aap}{A\&A}
\newcommand{\aaps}{A\&AS}
\newcommand{\aj}{AJ}
\newcommand{\mnras}{MNRAS}
\newcommand{\pasp}{PASP}
\newcommand{\spie}{SPIE}
\newcommand{\nodata}{\ensuremath{-}}

\begin{document}

\title[{\it Chandra} spectroscopy of $\beta$ Cru]{{\it Chandra}
  spectroscopy of the hot star $\beta$ Crucis and the discovery of a
  pre-main-sequence companion}

\author[D. Cohen et al.]{David H.\ Cohen,$^{1}$\thanks{E-mail: cohen@astro.swarthmore.edu} Michael A.\ Kuhn,$^{1}$\thanks{Current address: Penn State University, Department of Astronomy and Astrophysics, University Park, Pennsylvania 16802} Marc Gagn\'{e},$^{2}$ Eric L.\ N.\ Jensen,$^{1}$ \newauthor Nathan A.\ Miller$^{3}$ \\
  $^{1}$Swarthmore College, Department of Physics and Astronomy, Swarthmore, Pennsylvania 19081, USA\\
  $^{2}$West Chester University of Pennsylvania, Department of Geology and Astronomy, West Chester, Pennsylvania 19383, USA \\
  $^{3}$University of Wisconsin-Eau Claire, Department of Physics and
  Astronomy, Eau Claire, Wisconsin 54702, USA }

\maketitle

\label{firstpage}

\begin{abstract}

  In order to test the O star wind-shock scenario for X-ray production
  in less luminous stars with weaker winds, we made a pointed 74 ks
  observation of the nearby early B giant, \bcru\ (B0.5 III), with the
  \chandra\ High Energy Transmission Grating Spectrometer.  We find
  that the X-ray spectrum is quite soft, with a dominant thermal
  component near 3 million K, and that the emission lines are resolved
  but quite narrow, with half-widths of 150 \kms. The
  forbidden-to-intercombination line ratios of Ne\, {\sc ix} and Mg\,
  {\sc xi} indicate that the hot plasma is distributed in the wind,
  rather than confined near the photosphere. It is difficult to
  understand the X-ray data in the context of the standard wind-shock
  paradigm for OB stars, primarily because of the narrow lines, but
  also because of the high X-ray production efficiency.  A scenario in
  which the bulk of the outer wind is shock heated is broadly
  consistent with the data, but not very well motivated theoretically.
  It is possible that magnetic channeling could explain the X-ray
  properties, although no field has been detected on \bcru.  We
  detected periodic variability in the hard ($h\nu > 1$ keV) X-rays,
  modulated on the known optical period of 4.58 hours, which is the
  period of the primary \bcep\/ pulsation mode for this star. We also
  have detected, for the first time, an apparent companion to \bcru\/
  at a projected separation of 4\arcsec.  This companion was likely
  never seen in optical images because of the presumed very high
  contrast between it and \bcru\/ in the optical.  However, the
  brightness contrast in the X-ray is only 3:1, which is consistent
  with the companion being an X-ray active low-mass pre-main-sequence
  star.  The companion's X-ray spectrum is relatively hard and
  variable, as would be expected from a post T Tauri star.  The age of
  the \bcru\/ system (between 8 and 10 Myr) is consistent with this
  interpretation which, if correct, would add \bcru\/ to the roster of
  Lindroos binaries -- B stars with low-mass pre-main-sequence
  companions.

\end{abstract}


\section{Introduction} \label{sec:intro}

X-ray emission in normal O and early B stars is generally thought to
arise in shocked regions of hot plasma embedded in the fast,
radiation-driven stellar winds of these very luminous objects.
High-resolution X-ray spectroscopy of a handful of bright O stars has
basically confirmed this scenario \citep{Kahn2001, kco2003, Cohen2006,
  Leutenegger2006}.  The key diagnostics are line profiles, which are
Doppler broadened by the wind outflow, and the
forbidden-to-intercombination (\ftoi) line ratios in helium-like ions
of Mg, Si, and S, which are sensitive to the distance of the shocked
wind plasma from the photosphere.  Additionally, these O stars
typically show rather soft spectra and little X-ray variability, both
of which are consistent with the theoretical predictions of the
line-driving instability (LDI) scenario \citep{OCR1988} and in
contrast to observations of magnetically active coronal sources and
magnetically channeled wind sources.


\begin{center}

\begin{table*}
\begin{minipage}{180mm}
\caption{Properties of $\beta$ Cru}
\begin{tabular}{lcl}
\hline

MK Spectral Type & B0.5 III & \citet{hgs1969}.  \\
Distance (pc)  & $108 \pm 7$  & {\it Hipparcos} \citep{Perryman1997}. \\
Age (Myr) & $8$ to $11$ &\citet{Lindroos1985}; \citet{gzl1989}; \citet{Bertelli1994} and this work. \\
$\theta_{\rm LD}$ (mas) & $0.722 \pm 0.023$ & \citet{HB1974}. \\
$M$ ($\Msun$) & $16$ & \citet{Aerts1998}. \\
\Teff\/ (K) & $27,000 \pm 1000$ & \citet{Aerts1998}. \\
log $g$ (cm s$^{-2}$) & $3.6 \pm 0.1$; $3.8 \pm 0.1$ & \citet{Aerts1998}; Mass from \citet{Aerts1998} combined with radius. \\
$v \sin i$ (\kms) & $35$ & \citet{uf1982}. \\
$L$ ($\Lsun$) & $3.4 \times 10^4$ & \citet{Aerts1998}. \\
$R$ ($\Rsun$) & $8.4 \pm 0.6$ & From distance \citep{Perryman1997} and $\theta_{\rm LD}$ \citep{HB1974}. \\
Pulsation periods (hr) & 4.588, 4.028, 4.386, 6.805, 8.618 & \citet{Aerts1998,Cuypers2002}. \\
\mdot\/ (\Msunyr) &  $10^{-8}$ & Theoretical calculation, using \citet{Abbott1982}. \\
${\mdot}q$ (\Msunyr) &  $10^{-11}$ & Product of mass-loss rate and ionization fraction of  C\,{\sc iv} \citep{Prinja1989}. \\
\vinf\/ (km s$^{-1}$) & $2000$ & Theoretical calculation, using \citet{Abbott1982}. \\
\vinf\/ (km s$^{-1}$) & $420$ & Based on C\,{\sc iv} absorption feature blue edge \citep{Prinja1989}. \\

\hline
\end{tabular}
\label{tab:properties}
\end{minipage}
\end{table*}  

\end{center}

In order to investigate the applicability of the LDI wind-shock
scenario to early B stars, we have obtained a pointed \chandra\
grating observation of a normal early B star, $\beta$ Crucis (B0.5
III). The star has a radiation-driven wind but one that is much weaker
than those of the O stars that have been observed with \chandra\ and
\xmm.  Its mass-loss rate is at least two and more likely three orders
of magnitude lower than that of the O supergiant \zpup, for example.
Our goal is to explore how wind-shock X-ray emission changes as one
looks to stars with weaker winds, in hopes of shedding more light on
the wind-shock mechanism itself and also on the properties of early B
star winds in general. We will apply the same diagnostics that have
been used to analyze the X-ray spectra of O stars: line widths and
profiles, \ftoi\ ratios, temperature and abundance analysis from
global spectral modeling, and time-variability analysis.

There are two other properties of this particular star that make this
observation especially interesting: \bcru\ is a \bcep\/ variable,
which will allow us to explore the potential connection among
pulsation, winds, and X-ray emission; and the fact that there are
several low-mass pre-main-sequence stars in the vicinity of \bcru.
This second fact became potentially quite relevant when we
unexpectedly discovered a previously unknown X-ray source four arc
seconds from \bcru\ in the \chandra\/ data.

In the next section, we discuss the properties and environment of
\bcru.  In \S 3 we present the \chandra\ data.  In \S 4 we analyze the
spectral and time-variability properties of \bcru.  In \S 5 we perform
similar analyses of the newly discovered companion.  We discuss the
implications of the analyses of both \bcru\ and the companion as well
as summarize our conclusions in \S 6.

\section{The \bcru\/ system} \label{sec:the_star}

One of the nearest early-type stars to Earth, \bcru\ is located at a
distance of only 108 pc \citep{Perryman1997}\footnote{A recent
  re-reduction of the Hipparcos data
  \citep{vanLeeuwen2007a,vanLeeuwen2007b} results in a smaller
  distance of $85^{+8}_{-6}$ pc; or $88^{+8}_{-7}$ pc with the
  application of the Lutz-Kelker correction \citep{Maiz2008}.  This
  re-evaluation of the distance has a small effect on the results
  presented here.  It makes the radius inferred from the angular
  diameter correspondingly smaller and the gravity larger, as well as
  decreasing the X-ray luminosity inferred from the measured flux. In
  any case, parallax distance determinations may be affected by the
  spectroscopic binary companion, and a highly precise distance
  determination should attempt to correct for the small orbital
  angular displacement of \bcru\/ during the course of its 5 year
  orbit.}.  With a spectral type of B0.5, this makes \bcru\ -- also
known as Becrux, Mimosa, and HD 111123 -- extremely bright. In fact,
it is the 19th brightest star in the sky in the V band, and as a
prominent member of the Southern Cross (Crux), it appears in the flags
of at least five nations in the Southern Hemisphere, including
Australia, New Zealand, and Brazil.

The fundamental properties of \bcru\ have been extensively studied
and, in our opinion, quite well determined now, although binarity and
pulsation issues make some of this rather tricky.  In addition to the
Hipparcos distance, there is an optical interferometric measurement of
the star's angular diameter \citep{HB1974}, and thus a good
determination of its radius.  It is a spectroscopic binary
\citep{Heintz1957} with a period of 5 years \citep{Aerts1998}. Careful
analysis of the spectroscopic orbit, in conjunction with the
brightness contrast at 4430 \AA\ from interferometric measurements
\citep{Popper1968} and comparison to model atmospheres, enable
determinations of the mass, effective temperature, and luminosity of
\bcru\ \citep{Aerts1998}.  The log $g$ value and mass imply a radius
that is consistent with that derived from the angular diameter and
parallax distance, giving us additional confidence that the basic
stellar parameters are now quite well constrained. The projected
rotational velocity is low ($v \sin i=35$ \kms), but the inclination
is also likely to be low. \citet{Aerts1998} constrain the orbital
inclination to be between 15\degr\/ and 20\degr\/, and binary orbital
and equatorial planes are well aligned for binaries with $a \la 10$ AU
\citep{Hale1994}.  This implies $v_{\mathrm {rot}} \approx 120$ \kms\/
and combined with $\Rstar = 8.4 ~\Rsun$, the rotational period is
$P_{\mathrm {rot}} \approx 3\fd6$.  The star's properties are
summarized in Table \ref{tab:properties}.


With $\Teff = 27,000 \pm 1000$ K, \bcru\ lies near the hot edge of the
\bcep\/ pulsation strip in the HR diagram \citep{sh2005}.  Three
different non-radial pulsation modes with closely spaced periods
between 4.03 and 4.59 hours have been identified spectroscopically
\citep{Aerts1998}.  Modest radial velocity variations are seen at the
level of a few \kms\/ overall, with some individual pulsation
components having somewhat larger amplitudes \citep{Aerts1998}. The
three modes are identified with azimuthal wavenumbers $\ell = 1$, 3,
and 4, respectively \citep{Aerts1998,Cuypers2002}. The observed
photometric variability is also modest, with amplitudes of a few
hundredths of a magnitude seen only in the primary pulsation mode
\citep{Cuypers1983}.  More recent, intensive photometric monitoring
with the star tracker aboard the WIRE satellite has identified the
three spectroscopic periods and found two additional very low
amplitude (less than a millimagnitude), somewhat lower frequency
components \citep{Cuypers2002}.

The wind properties of \bcru\ are not very well known, as is often the
case for non-supergiant B stars.  The standard theory of line-driven
winds \citep{cak1975} (CAK) enables one to predict the mass-loss rate
and terminal velocity of a wind given a line list and stellar
properties.  There has not been much recent work on applying CAK
theory to B stars, as the problem is actually significantly harder
than for O stars, because B star winds have many fewer constraints
from data and the ionization/excitation conditions are more difficult
to calculate accurately.  With those caveats, however, we have used
the CAK parameters from \citet{Abbott1982} and calculated the
mass-loss rate and terminal velocity using the stellar parameters in
Table \ref{tab:properties} and the formalism described by
\citet{Kudritzki1989}.  The mass-loss rate is predicted to be about
$10^{-8}$ \Msunyr\ and the terminal velocity, $\vinf \approx 2000$
\kms.  There is some uncertainty based on the assumed ionization
balance of helium, as well as the uncertainty in the stellar
properties (gravity, effective temperature, luminosity), and whatever
systematic errors exist in the line list. The wind models of
\citet{Vink2000} also predict a mass-loss rate close to $10^{-8}$
\Msunyr\ for \bcru.

Interestingly, \iue\ observations show very weak wind signatures - in
Si\,{\sc iv} and C\,{\sc iv}, with no wind signature at all seen in
Si\,{\sc iii}.  The C\,{\sc iv} doublet near 1550 \AA\/ has a blue
edge velocity of only 420 \kms, while the blue edge velocity of the
Si\,{\sc iv} line is slightly smaller.  The products of the mass-loss
rate and the ionization fraction of the relevant ion are roughly
$10^{-11}$ \Msunyr\ for both of these species \citep{Prinja1989}.
Either the wind is much weaker than CAK theory (using Abbott's
parameters) predicts or the wind has a very unusual ionization
structure.  We will come back to this point in \S
\ref{sec:discussion}.

\bcru\ is most likely a member of the Lower Centaurus Crux (LCC)
subgroup of the Sco-Cen OB association.  Although a classical member
of LCC \citep[e.g,][]{Blaauw1946}, \bcru\ was not selected as an LCC
member by \citet{deZeeuw1999} in their analysis of Hipparcos data.
However, the timespan of the Hipparcos proper motion measurement (4
years) is similar to the 5-yr orbital period of the spectroscopic
binary companion. The rate of orbital angular displacement of \bcru\
is similar to its proper motion, which likely influenced the Hipparcos
proper-motion measurement.  For this reason, \citet{deZeeuw1999}
suspect that \bcru\ is an LCC member even though it is not formally
selected as one by their criteria, and indeed \citet{Hoogerwerf2000}
find that \bcru\ is an LCC member based on longer-time-baseline proper
motion measurements.

The age of \bcru\ is also roughly consistent with LCC membership,
given the uncertainty about the age(s) of the LCC itself.  The age of
LCC has been variously found to be from 10--12 Myr \citep{gzl1989} to
17--23 Myr \citep{Mamajek2002}, and more recent work has found
evidence for an age spread and spatial substructure within the region
\citep{PreibischMamajek2007}.  \bcru\ has been claimed to have an
evolutionary age of 8 Myr \citep{Lindroos1985}, though the distance
was much more uncertain when this work was done.  Using our adopted
stellar parameters (Table \ref{tab:properties}), \bcru\ has an age of
8--10 Myr when plotted on the evolutionary tracks of
\citet{Bertelli1994}.  As such, \bcru\/ has evolved off the main
sequence, but not very far.  Its spectroscopic binary companion is a
B2 star that is still on the main sequence \citep{Aerts1998}, and the
projected X-ray companion discussed below, if it is a physical
companion, is likely to be a pre--main-sequence (PMS) star.  There are
two other purported wide visual companions (\bcru\ B and C) listed in
the Washington Double Star catalog \citep{wd1997}, with separations of
44\arcsec\ and 370\arcsec, although they are almost certainly not
physical companions of \bcru\/ \citep{Lindroos1985}.

A small group of X-ray bright pre--main-sequence stars within about a
degree of \bcru\ was discovered by \rosat\ \citep{pf1996}.  These
stars are likely part of the huge low-mass PMS population of the LCC
\citep{fl1997}.  They have recently been studied in depth
\citep{Alcala2002}, and four of the six have been shown to be likely
post-T-Tauri stars (and one of those is a triple system).  They have
${\mathrm H{\alpha}}$ in emission, strong Li 6708 \AA\/ absorption,
and radial velocities consistent with LCC membership.  Assuming that
they are at the same distance as \bcru, their ages from PMS
evolutionary tracks are 4--10 Myr.  This is somewhat younger than LCC
ages derived from the more massive members, but this is consistent
with a larger pattern of systematic age differences for LCC members of
different masses and may indicate systematic errors in the
evolutionary tracks \citep{PreibischMamajek2007}.

\section{The \chandra\/ data} \label{sec:data}

We obtained the data we report on in this paper in a single long
pointing on 28 May 2002, using the \chandra\/ Advanced CCD Imaging
Spectrometer in spectroscopy mode with the High Energy Transmission
Grating Spectrometer (ACIS-S/HETGS configuration)
\citep{Canizares2005}. The ACIS detector images the dispersed spectra
from two grating components of the HETGS, the medium energy grating
(MEG) and the high energy grating (HEG). The mirror, grating, and
detector combination has significant response on a wavelength range
extending roughly from 2 \AA\/ to 40 \AA, though with a total
(including both negative and positive orders) effective area of only
about 10 cm$^2$.  The spectral resolution exceeds
$\lambda/\Delta{\lambda} = 1000$ at the long-wavelength end of each
grating's spectral range, and decreases toward shorter wavelengths.
The spatial resolution of \chandra\/ is excellent, with the core of
the point spread function having a full width of less than half an
arcsecond.  Properties and characteristics of the telescope and
detector are available via the Proposers' Observatory Guide
\citep{CXC2007}.

We reran the pipeline reduction tasks in the the standard \chandra\/
Interactive Analysis of Observations (CIAO) software, version 3.3,
using the calibration database, CALDB 3.2, and extracted the zeroth
order spectrum and the dispersed first order MEG and HEG spectra of
both \bcru\/ and the newly discovered companion.  The zeroth order
spectrum of a source in the \chandra\ ACIS-S/HETGS is an image created
by source photons that pass directly through the transmission gratings
without being diffracted. The ACIS CCD detectors themselves have some
inherent energy discrimination so a low-resolution spectrum can be
produced from these zeroth-order counts.


\begin{figure}
\includegraphics[angle=0,width=80mm]{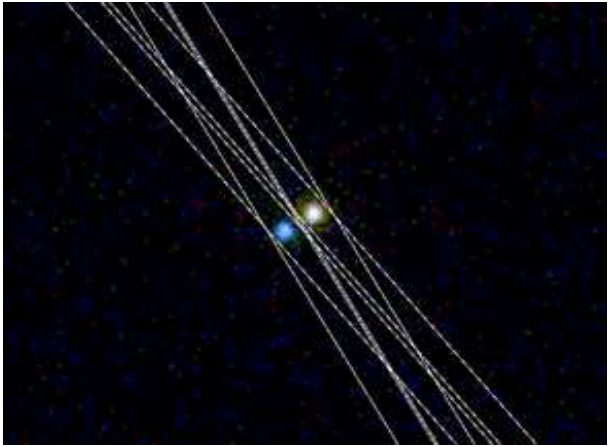}
\caption{Image of the center of the ACIS detector, showing the zeroth
  order spectra of \bcru\ and its newly discovered companion.  North
  is up and east is left.  \bcru\/ is the source to the northwest.
  Its position in the ACIS detector is coincident with the optical
  position of \bcru\/ to within the positional accuracy of \chandra.
  The companion is to the southeast, at a separation of 4.0\arcsec\/
  and with a position angle relative to \bcru\/ of $120^{\circ}$.  The
  extraction regions for the MEG and HEG (both positive and negative
  orders) are indicated for both sources.  It can be seen in this
  image that the position angle is nearly perpendicular to the
  dispersion directions (for both the MEG and HEG) and thus that the
  dispersed spectra of the two sources can be separated.  In the
  on-line color version of this figure, hard counts (h$\nu>1$ keV) are
  blue, soft counts (h$\nu<0.5$ keV) are red, and counts with
  intermediate energies are green.  It can easily be seen in the color
  image that the X-ray emission of the companion is dramatically
  harder than that of \bcru.}
\label{fig:finding_chart}
\end{figure}


The zeroth order spectrum of each source has very little pileup ($\sim
7$\% for \bcru), and the grating spectra have none.  The HEG spectra
and the higher order MEG spectra have almost no source counts, so all
of the analysis of dispersed spectra we report on here involves just
the MEG $\pm$1 orders. There were typically five to ten pixels across
each emission line, and few lines have even 100 counts per line.  The
random errors on the source count rates are dominated by Poisson
photon-counting noise. Thus even the strongest lines have
signal-to-noise ratios per pixel of only several. The background
(instrumental and astrophysical) is negligible.  We made
observation-specific response matrix files ({\it rmfs}; encoding the
energy-dependent spectral response) and grating ancillary response
files ({\it garfs}; containing information about the
wavelength-dependent effective area) and performed our spectral
analysis in both XSPEC 11.3.1 and ISIS 1.4.8.  We performed
time-variability analysis on both binned light curves and unbinned
photon event tables using custom-written codes.


The observation revealed only a small handful of other possible
sources in the field, which we identified using the CIAO task {\it
  celldetect} with the threshold parameter set to a value of 3. We
extracted background-subtracted source counts from three weak sources
identified with this procedure.  Their signal-to-noise ratios are each
only slightly above 3. The only other bright source in the field of
view is one just 4\farcs0 to the southeast of \bcru\ at a position
angle of 120 degrees, which is clearly separated from \bcru, as can be
seen in Fig.\ \ref{fig:finding_chart}. The position angle uncertainty
is less than a degree based on the formal errors in the centroids of
the two sources, and the uncertainty in the separation is less than
0\farcs1.  We present the X-ray properties of this source in
\S\ref{sec:companion} and argue in \S\ref{sec:discussion} that it is
most likely a low-mass PMS star in orbit around \bcru. The B star that
dominates the optical light of the \bcru\/ system and which is the
main subject of this paper should formally be referred to as \bcru\ A
(with spectroscopic binary components 1 and 2).  The companion at a
projected separation of 4\arcsec\/ will formally be known as \bcru\ D,
since designations B and C have already been used for the purported
companions listed in the Washington Double Star Catalog, mentioned in
the previous section. Throughout this paper, we will refer to \bcru\ A
and D simply as ``\bcru'' and ``the companion.''


\begin{table*}
\begin{minipage}{160mm}
\caption{Point sources in the field}

\begin{tabular}{lcccc}
  \hline
  Name(s) &  RA\footnote{The positions listed here are not corrected for any errors in the aim point of the telescope.  There is an offset of about 0\farcs7 between the position of \bcru\/ in the \chandra\/ observation and its known optical position. This likely represents a systematic offset for all the source positions listed here.  The statistical uncertainties in the source positions, based on random errors in the centroiding of their images in the ACIS detector, are in all cases $\la 0\farcs1$.} & Dec.  & total counts\footnote{Zeroth order spectrum, background subtracted using several regions near the source to sample the background.} & X-ray flux\footnote{On the range $0.5~{\rm keV} < {\rm h}\nu < 8~{\rm keV}$ and, for the weak sources, based on fits to the zeroth order spectra using a power law model with interstellar absorption.  For the brighter sources (\bcru\/ and its newly discovered companion) the flux is based on two-temperature APEC model fits discussed in \S4.1 and \S5.1.} \\
  & (J2000) & (J2000) & & ($10^{-14}$ ergs s$^{-1}$ cm$^{-2}$) \\
  \hline
  $\beta$ Cru A; Mimosa; Becrux; HD 111123  & $12~47~43.35$  & $-59~41~19.2$ & $3803 \pm 12$ & $186 \pm 1$ \\
  $\beta$ Cru D; CXOU J124743.8-594121  & $12~47~43.80$  & $-59~41~21.3$ & $1228 \pm 12$ & $57.3 \pm .6$ \\
  CXOU J124752.5-594345  &  $12~47~52.53$  & $-59~43~45.8$ & $25.9 \pm 8.6$ & $1.58 \pm .52$ \\
  CXOU J124823.9-593611  &  $12~48~23.91$  & $-59~36~11.8$ & $47.4 \pm 10.4$ & $3.32 \pm .73$ \\
  CXOU J124833.2-593736  &  $12~48~33.22$  & $-59~37~36.8$ & $29.6 \pm 9.8$  & $2.23 \pm .74$ \\
  \hline
\end{tabular}
\label{tab:other_sources}
\end{minipage}
\end{table*}

We summarize the properties of all the point sources on the ACIS chips
in Tab.\ \ref{tab:other_sources}. We note that the other three sources
have X-ray fluxes about two orders of magnitude below those of \bcru\/
and the newly discovered companion.  They show no X-ray time
variability, and their spectra have mean energies of roughly 2 keV and
are fit by absorbed power laws.  These properties, along with the fact
that they have no counterparts in the 2MASS point source catalog,
indicate that they are likely AGN. We did not detect any of the
pre-main-sequence stars seen in the \rosat\/ observation because none
of them fell on the ACIS CCDs.  We also note that there is no X-ray
source detected at the location of the wide visual companion located
44\arcsec\/ from \bcru, \bcru\ C, which is no longer thought to be
physically associated with \bcru\ \citep{Lindroos1985}.  We place a
$1\sigma$ upper limit of 7 source counts in an extraction region at
the position of this companion.  This limit corresponds to an X-ray
flux more than two orders of magnitude below that of \bcru\/ or its
X-ray bright companion.

The position angle of the newly discovered companion is oriented
almost exactly perpendicular to the dispersion direction of the MEG
for this particular observation.  This can be seen in Fig.\
\ref{fig:finding_chart}, where parts of the rectangular extraction
regions for the dispersed spectra are indicated.  We were thus able to
cleanly extract not just the zeroth order counts for both sources but
also the MEG and HEG spectra for both sources.  We visually inspected
histograms of count rates versus pixel in the cross-dispersion
direction at the locations of several lines and verified that there
was no contamination of one source's spectrum by the other's. We
analyze these first-order spectra as well as the zeroth-order spectra
and the timing information for both sources in \S\ref{sec:primary} and
\S\ref{sec:companion}.

\section{Analysis of \bcru} \label{sec:primary}

\subsection{Spectral Analysis} \label{sec:spectral_analysis_primary}

\begin{figure}
\includegraphics[angle=90,width=80mm]{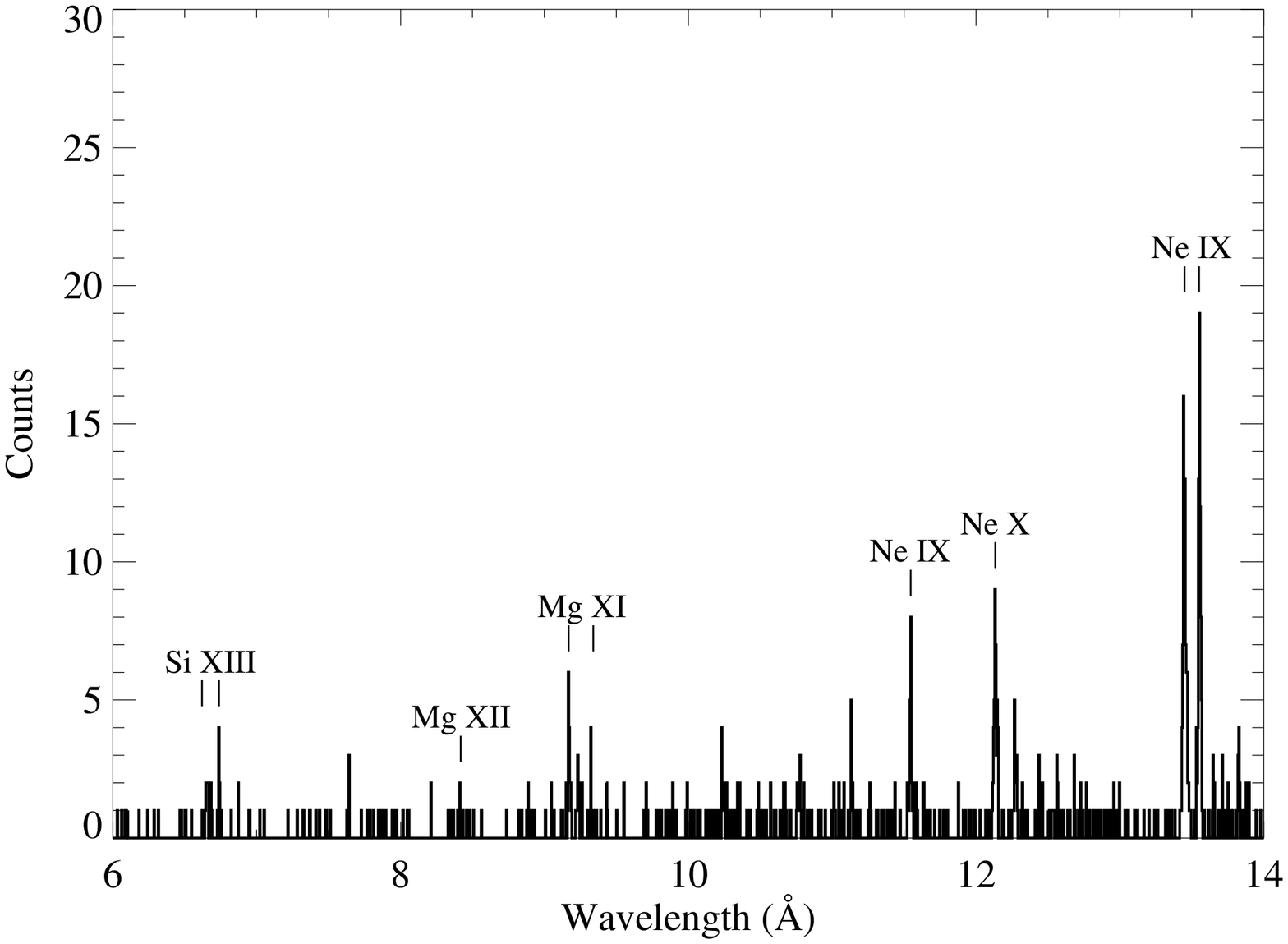}
\includegraphics[angle=90,width=80mm]{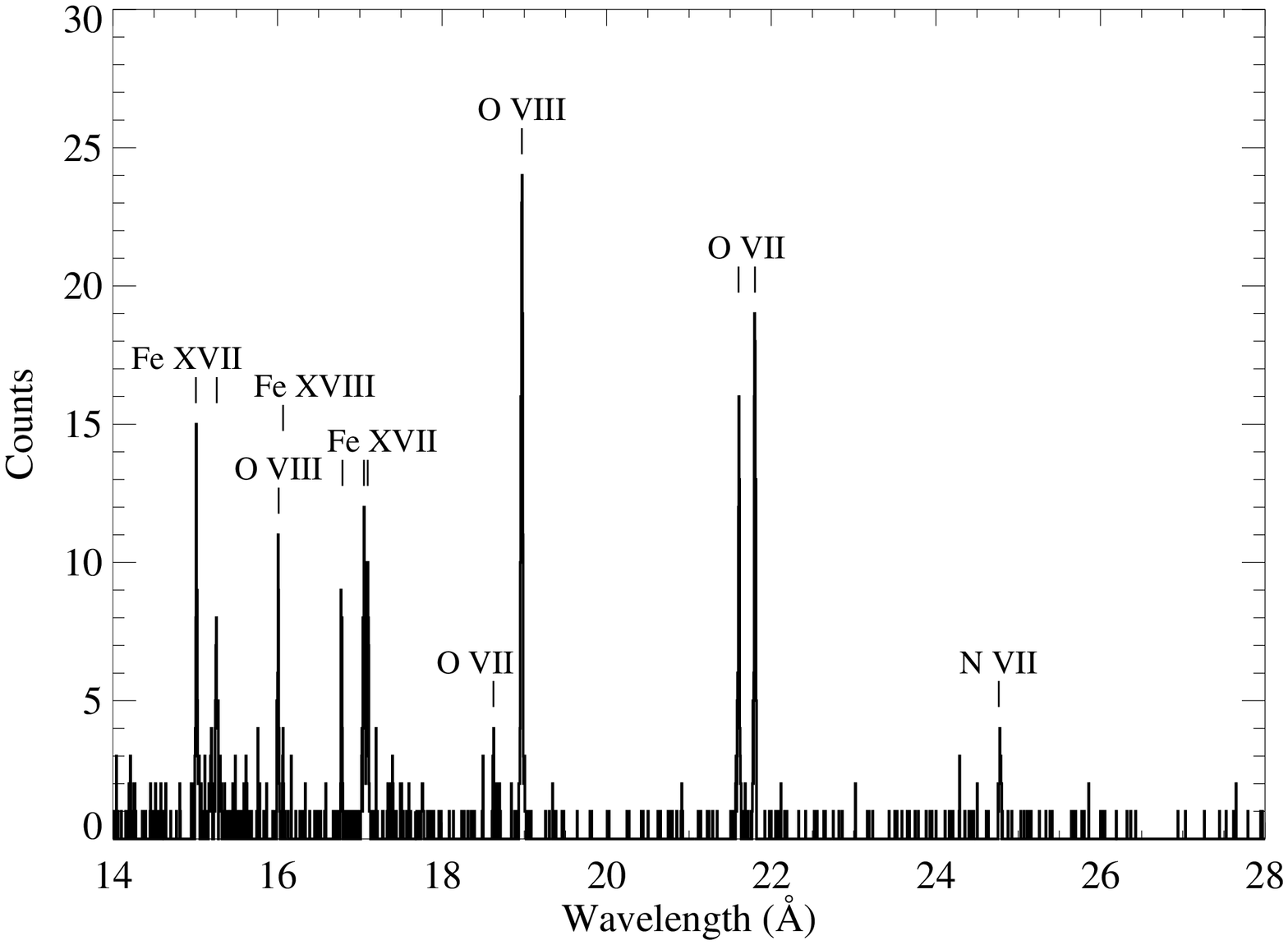}
\caption{The MEG spectrum of \bcru, with negative and positive first
  orders coadded and emission lines identified by ion. Bins are 5
  m\AA\ wide, and each emission line spans several bins. We show a
  smoothed version of these same data in Fig.\ \ref{fig:apec_meg_fit}.
}
\label{fig:atlas}
\end{figure}

The dispersed spectrum of \bcru\/ is very soft and relatively weak. We
show the co-added negative and positive first-order MEG spectra in
Fig.\ \ref{fig:atlas}, with the strong lines labeled.  The softness is
apparent, for example, in the relative weakness of the Ne\,{\sc x}
Lyman-$\alpha$ line at 12.13 \AA\/ compared to the Ne\,{\sc ix}
He$_{\alpha}$ complex near 13.5 \AA, and indeed the lack of any strong
lines at wavelengths shorter than 12 \AA.



\subsubsection{Global thermal spectral modeling}

We characterized the temperature and abundance distributions in the
hot plasma on \bcru\/ by fitting Astrophysical Plasma Emission Code
(APEC) thermal, optically thin, equilibrium plasma spectral emission
models \citep{Smith2001} to both the zeroth and first order spectra.
We included a model for pileup in the fit to the zeroth order spectrum
\citep{Davis2001}. For the dispersed MEG spectrum, we adaptively
smoothed the data for the fitting in ISIS.  This effectively combines
bins with very few counts to improve the statistics in the continuum.
We also modified the vAPEC (``v'' for variable abundances)
implementation in ISIS to allow the helium-like
forbidden-to-intercombination line ratios to be free parameters (the
so-called ${\mathcal G}$ parameter, which is the ratio of the
forbidden plus intercombination line fluxes to the flux in the
resonance line, ${\mathcal G} \equiv \frac{f+i}{r}$, was also allowed
to be a free parameter).  We discuss these line complexes and their
diagnostic power later in this subsection.  The inclusion of these
altered line ratios in the global modeling was intended at this stage
simply to prevent these particular lines from influencing the fits in
ways that are not physically meaningful.

The number of free parameters in a global thermal model like APEC can
quickly proliferate.  We wanted to allow for a non-isothermal plasma,
non-solar abundances, and thermal and turbulent line broadening (in
addition to the non-standard forbidden-to-intercombination line
ratios).  However, we were careful to introduce new model parameters
only when they were justified. So, for example, we only allowed
non-solar abundances for species that have several emission lines in
the MEG spectrum.  And we found we could not rely solely on the
automated fitting procedures in XSPEC and ISIS because information
about specific physical parameters is sometimes contained in only a
small portion of the spectrum, so that it has negligible effect on the
global fit statistic.

We thus followed an iterative procedure in which we first fit the
zeroth-order spectrum with a solar-abundance, two-temperature APEC
model.  We were able to achieve a good fit with this simple type of
model, but when we compared the best fit model to the dispersed
spectrum, systematic deviations between the model and data were
apparent.  Simply varying all the free model parameters to minimize
the fit statistic (we used both the Cash C statistic \citep{Cash1979}
and $\chi^2$, periodically comparing the results for the two
statistics) was not productive because, for example, the continuum in
the MEG spectrum contains many more bins than do the lines, but the
lines generally contain more independent information. And for some
quantities, like individual abundances, only a small portion of the
spectrum shows any dependence on that parameter, so we would hold
other model parameters fixed and fit specific parameters on a
restricted subset of the data. Once a fit was achieved, we would hold
that parameter fixed and free others and refit the data as a whole.



\begin{figure}
\includegraphics[angle=90,width=80mm]{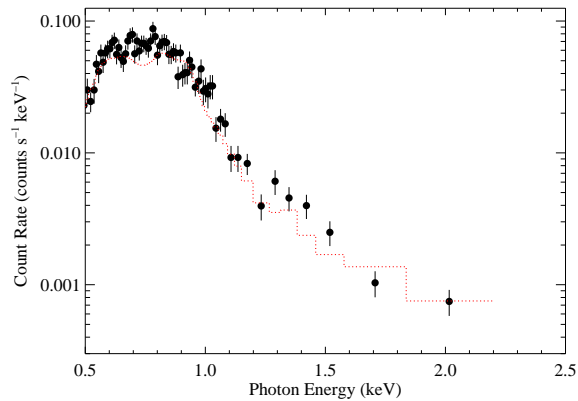}
\caption{The best-fit two-temperature, variable abundance thermal
  spectral (APEC) model (dotted line), corrected for pileup,
  superimposed on the zeroth-order spectrum of \bcru. The model shown
  is the best-fit model fit to both the zeroth and first order
  spectra, with parameters listed in Tab.\ \ref{tab:apec_fit}. It
  seems likely that some of the systematic deviations in the softer
  channels are due to optical contamination, as discussed in \S4.1.1.}
\label{fig:apec_ccd_fit}
\end{figure}

\begin{figure*}
\includegraphics[angle=0,width=180mm]{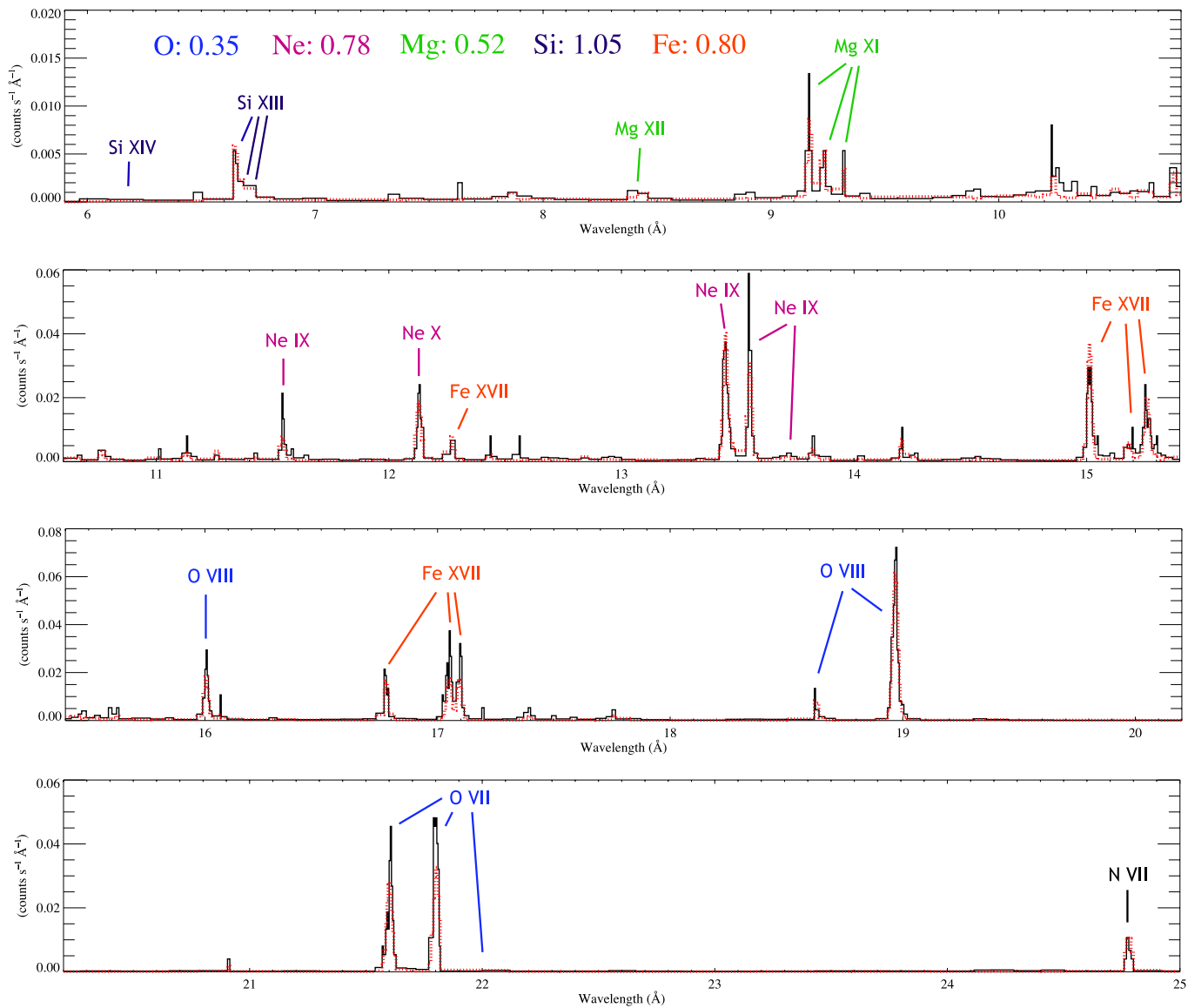}
\caption{The best-fit two-temperature, variable abundance thermal
  spectral (vAPEC) model (red, dotted line; parameters listed in Tab.\
  \ref{tab:apec_fit}; same model as shown in Fig.\
  \ref{fig:apec_ccd_fit}), superimposed on the adaptively smoothed
  co-added MEG first-order spectrum of \bcru.  Emission lines are
  labeled and color coded according to element, with the abundances of
  each element listed at the top of the figure.  Some non-detected
  lines are labeled, including the forbidden line of O\, {\sc vii}
  near 22.1 \AA\ and the Lyman-alpha line of Si\, {\sc xiv} near 6.2
  \AA. }
\label{fig:apec_meg_fit}
\end{figure*}

Ultimately, we were able to find a model that provided a good fit to
the combined data, both zeroth-order and dispersed spectra.  The model
parameters are listed in Tab.\ \ref{tab:apec_fit}, and the model is
plotted along with the zeroth order spectrum in Fig.\
\ref{fig:apec_ccd_fit} and the adaptively smoothed first order MEG
spectrum in Fig.\ \ref{fig:apec_meg_fit}. There are some systematic
deviations between the model and the data, but overall the fit quality
is good, with $\chi^2_{\nu} = 1.09$ for the combined data.  The formal
confidence limits on the fit parameters are unrealistically narrow, at
least partly because the statistical error cannot be fully accounted
for. For example, the abundance determinations are each made on narrow
wavelength intervals and thus the temperature cannot reliably be
allowed to vary along with the abundances in the confidence limit
testing procedure. We return to this point below when we discuss the
abundance determinations.  The overall result, that a thermal model
with a dominant temperature near 3 million K and an emission measure
of $\sim 3.5 \times 10^{53}$ cm$^{-3}$ and abundances that overall are
slightly sub-solar is required to fit the data, is quite robust.

The model we present provides a better fit for the first-order MEG
spectrum, which has 443 bins when we adaptively smooth it.  The model
gives $\chi_{\nu}^2 = 1.00$ for the first order MEG spectrum alone,
while it gives $\chi_{\nu}^2 = 1.42$ for the zeroth order spectrum
alone, which has 90 bins.  We suspect the relatively poor agreement
between the CCD spectrum and the MEG spectrum stems from the small,
but non-zero, transmission of red optical light through the MEG and
HEG gratings and through the ACIS optical blocking filter
\citep{Wolk2002}.  Being so bright, \bcru\/ is more susceptible to
this effect than nearly any other object.  We examined the \chandra\/
ACIS spectrum of Vega, which is very bright in the optical and X-ray
dark. Its CCD spectrum, which is presumably due entirely to the
optical contamination, looks very similar to the excess above the
model seen in the \bcru\/ spectrum. The optical contamination is
particularly noticeable in the $0.22 - 0.5$ keV ACIS-S band, which we
have chosen not to model. There appears to be some contamination up to
$\sim 0.7$ keV, however.



In Fig.\ \ref{fig:apec_meg_fit}, where we show our global best-fit
model superimposed on the adaptively smoothed MEG spectrum, the
continuum is seen to be well reproduced and the strengths of all but
the very weakest lines are reproduced to well within a factor of two.
It is interesting that no more than two temperature components are
needed to fit the data; the continuum shape and line strengths are
simultaneously explained by a strong 2.5 million K component and a
much weaker 6.5 million K component.

The constraints on abundances from X-ray emission line spectra are
entangled with temperature effects.  So, it is important to have line
strength measurements from multiple ionization stages. This is
strictly only true in the MEG spectrum for O and Ne, but important
lines of Mg, Si, and Fe from non-dominant ionization stages would
appear in the spectrum if they were strong, so the non-detections of
lines of Mg\, {\sc xii}, Si\, {\sc xiv}, and Fe\, {\sc xviii} provide
information about the abundances of these elements.  These five
elements are the only ones for which we report abundances derived from
the model fitting (see Tab.\ \ref{tab:apec_fit}).  We allowed the
nitrogen abundance to be a free parameter of the fit as well, but do
not report its value, as it is constrained by only one detected line
in the data.  The uncertainties on these abundances are difficult to
reliably quantify (because of the interplay with the temperature
distribution) but they are probably less than a factor of two, based
on our experience varying these parameter values by hand. They are
certainly somewhat larger than the formal confidence limits listed in
Tab.\ \ref{tab:apec_fit}, but even given that, there is an overall
trend of slightly subsolar abundances, consistent with a value perhaps
somewhat above half solar.  Only the oxygen abundance seems to be
robustly and significantly sub-solar. We note, however, that in
general, the abundances we find are significantly closer to solar than
those reported by \citet{zp2007}.


\begin{table}
  \caption{Best-fit parameters for the two-temperature APEC
    thermal equilibrium spectral model}
\begin{tabular}{ccc}
  \hline
parameter & value & 90\% confidence limits \\

  \hline

$T_{\mathrm 1}$ (K) & $2.5 \times 10^6$ & $2.4:2.6  \times 10^6$ \\
 ${\mathcal EM}_{\mathrm 1}$ (cm$^{-3}$) & $3.3 \times 10^{53}$ & $3.1:3.5 \times 10^{53}$ \\
$T_{\mathrm 2}$ (K) & $6.5 \times 10^6$  & $6.2:6.8  \times 10^6$\\
 ${\mathcal EM}_{\mathrm 2}$ (cm$^{-3}$) & $2.8  \times 10^{52}$ & $2.3:3.3 \times 10^{52}$ \\
$v_{\mathrm {turb}}$ (\kms)  & 180 & 90:191 \\
O & 0.35 & 0.30:0.40 \\ 
Ne & 0.78 & 0.70:0.99 \\ 
Mg & 0.52 & 0.29:0.62 \\ 
Si & 1.05 & 0.68:1.56 \\ 
Fe & 0.80 & 0.70:0.90 \\ 
\hline
\end{tabular}

\medskip The abundances are the (linear) fraction of the solar
abundance, according to \citet{ag1989}.
\label{tab:apec_fit}
\end{table}


In all fits we included ISM absorption assuming a hydrogen column
density of $3.5 \times 10^{19}$ cm$^{-2}$.  This value is consistent with
the very low $E(B-V) = 0.002 \pm .011$ \citep{nd2005} and is the value
determined for the nearby star HD 110956 \citep{Fruscione1994}.  This
low ISM column produces very little attenuation, with only a few
percent of the flux absorbed even at the longest \chandra\/
wavelengths.  The overall X-ray flux of \bcru\ between $0.5$ keV and 8
keV is $1.86 \pm .01 \times 10^{-12}$ ergs s$^{-1}$ cm$^{-2}$,
implying $\lx = 2.6 \times 10^{30}$ ergs s$^{-1}$, which corresponds
to $\log \lx/L_{\rm Bol} = -7.72$.

\subsubsection{Individual emission lines: strengths and widths}

Even given the extreme softness of the dispersed spectrum shown in
Figs.\ \ref{fig:atlas}, \ref{fig:apec_ccd_fit}, and
\ref{fig:apec_meg_fit}, the most striking thing about these data is
the narrowness of the X-ray emission lines.  Broad emission lines in
grating spectra of hot stars are the hallmark of wind-shock X-ray
emission \citep{Kahn2001,Cassinelli2001,kco2003,Cohen2006}.  While the
stellar wind of \bcru\/ is weaker than that of O stars, it is expected
that the same wind-shock mechanism that operates in O stars is also
responsible for the X-ray emission in these slightly later-type early
B stars \citep{CCM1997}.  The rather narrow X-ray emission lines we
see in the MEG spectrum of \bcru\/ would thus seem to pose a severe
challenge to the application of the wind-shock scenario to this early
B star.  We will quantify these line widths here, and discuss their
implications in \S\ref{sec:discussion}.


\begin{table*}
  \begin{minipage}{175mm}
  \caption{Emission Lines in the HETGS/MEG Spectrum}
  \label{symbols}
  \begin{tabular}{lcccccc}
    \hline
    ion & $\lambda_{{\rm lab}}$ \footnote{
      Lab wavelength taken from the Astrophysical Plasma Emission Database (APED) \citep{Smith2001}. For closely spaced doublets 
      (e.g. Lyman-alpha lines) an emissivity-weighted mean wavelength
      is reported. The resonance (r), intercombination (i), and forbidden (f) lines of Si, Mg, Ne, and O are indicated. 
    }
    & $\lambda_{{\rm obs}}$ \footnote{If no value is given, then this parameter was held constant at the laboratory value when the fit was performed.}  & shift & half-width\footnote{For closely spaced lines, such as the He-like complexes, we tied the line widths together but treated the common line width as a free parameter.  This is indicated by identical half-width values and formal uncertainties listed on just the line with the shortest wavelength.}
    & normalization & corrected flux \footnote{
      Corrected for an ISM column density of N$_H=3.5\times10^{19}$ cm$^{-2}$ using 
      \citet{MM1983} photoelectric cross sections as 
      implemented in the XSPEC 
      model \emph{wabs}.
    }
    \\
    & (\AA) & (\AA) & (m\AA) & (m\AA) & (10$^{-6}$ ph s$^{-1}$ cm$^{-2}$) 
    & (10$^{-15}$ erg s$^{-1}$ cm$^{-2}$)   \\

    \hline

    Si\, {\sc xiii}& 6.6479 (r) & \nodata & \nodata & $1.7^{+3.5}_{-1.7}$
    & $1.06^{+0.41}_{-0.34}$& $3.15^{+1.22}_{-1.01}$  \\

    Si\, {\sc xiii}& 6.6882 (i)& \nodata & \nodata & $1.7$ & $0.31^{+0.26}_{-0.17}$ & $0.93^{+0.78}_{-0.51}$ \\

    Si\, {\sc xiii}& 6.7403 (f) & \nodata & \nodata & $1.7$ & $0.77^{+0.33}_{-0.26}$ & $2.26^{+0.97}_{-0.76}$ \\

    Mg\, {\sc xi}& 9.1687 (r)& \nodata & \nodata & $1.5^{+4.2}_{-1.5}$ & $2.21^{+0.63}_{-0.52}$ & $4.81^{+1.37}_{-1.13}$ \\

    Mg\, {\sc xi}& 9.2312 (i) & \nodata & \nodata  & $1.5$ & $1.25^{+0.50}_{-0.40}$ & $2.70^{+1.08}_{-0.86}$ \\

    Mg\, {\sc xi}& 9.3143 (f) & \nodata & \nodata & $1.5$ & $0.82^{+0.46}_{-0.30}$ & $1.56^{+1.04}_{-0.71}$ \\

    Ne\, {\sc ix}& 11.5440 & \nodata & \nodata & $1.3^{+12.8}_{-1.3}$ & $6.0^{+1.3}_{-1.4}$ &  $10.5^{+2.2}_{-2.5}$ \\

    Ne\, {\sc x}& 12.1339 & $12.1331\pm0.0019$& $-0.8\pm1.9$ & $6.3^{+2.1}_{-2.3}$ & $13.0^{+2.6}_{-1.9}$ & $21.5^{+4.3}_{-3.1}$ \\ 
  
    Ne\, {\sc ix}& 13.4473 (r) & \nodata & \nodata & $7.0\pm1.3$ & $39.9^{+4.0}_{-5.2}$ & $59.6^{+6.0}_{-6.6}$ \\

    Ne\, {\sc ix}& 13.5531 (i) & \nodata & \nodata & $7.0$ & $37.4_{-4.9}^{+4.2}$& $55.5_{-7.3}^{+6.1}$ \\

    Ne\, {\sc ix}& 13.6990 (f)& \nodata & \nodata & $7.0$ & $3.4_{-1.5}^{+1.7}$ & $4.9_{-2.2}^{+2.5}$ \\

    Fe\, {\sc xvii}& 15.0140 & $15.0146^{+0.0021}_{-0.0020}$ & $0.6^{+2.1}_{-2.0}$
                   & $10.0_{-2.0}^{+2.6}$ & $60.6_{-11.1}^{+5.7}$ & $81.1_{-14.9}^{+7.6}$ \\

    Fe\, {\sc xvii}& 15.2610 & \nodata & \nodata & $11.7^{+3.7}_{-3.0}$ & $28.1^{+5.2}_{-4.6}$ & $37.1^{+6.9}_{-6.1}$ \\

    O\, {\sc viii}& 16.0059 & $16.0053^{+0.0017}_{-0.0013}$& $-0.6^{+1.7}_{-1.3}$ & $0.9_{-0.8}^{+5.1}$ & $32.2_{-4.5}^{+6.1}$ & $40.5_{-5.7}^{+7.7}$ \\

    Fe\, {\sc xviii}& 16.0710 & \nodata & \nodata & $2.6_{-1.6}^{+4.4}$ & $8.2_{-2.5}^{+2.9}$ & $10.3_{-3.2}^{+3.9}$ \\

    Fe\, {\sc xvii}& 16.7800 & $16.7756^{+0.0044}_{-0.0010}$ & $-4.4^{+4.4}_{-1.0}$ & $0.8_{-0.8}^{+3.1}$ & $31.8_{-5.7}^{+5.9}$ &  $38.3_{-6.9}^{+7.1}$ \\

    Fe\, {\sc xvii}& 17.0510 & $17.0477_{-0.0017}^{+0.0020}$ & $-3.3_{-1.7}^{+2.0}$ & $8.8^{+1.5}_{-1.7}$ & $67.3^{+6.4}_{-5.8}$ & $78.4^{+9.0}_{-5.5}$ \\

    Fe\, {\sc xvii}& 17.0960 & $17.0938_{-0.0016}^{+0.0022}$ & $-2.2_{-1.6}^{+2.2}$ & $8.8$ & $53.2^{+6.3}_{-6.5}$ & $62.9^{+7.4}_{-7.7}$ \\

    O\, {\sc vii}& 18.6270 & \nodata & \nodata & $3.9^{+12.4}_{-3.9}$ & $21.1_{-5.9}^{+8.3}$ & $22.9_{-6.4}^{+9.0}$ \\

    O\, {\sc viii}& 18.9689 & $18.9671_{-0.0011}^{+0.0014}$ & $-1.8_{-1.1}^{+1.4}$ & $10.4\pm1.3$ & $260_{-24}^{+21}$ &  $278_{-26}^{+22}$ \\

    O\, {\sc vii}& $21.6015$ (r) & $21.6016^{+0.0018}_{-0.0017}$ & $-0.1^{+1.8}_{-1.7}$ & $11.0\pm1.8$ & $295\pm34$ & $279\pm30$ \\

    O\, {\sc vii}& $21.8038$ (i) & $21.7982^{+0.0012}_{-0.0009}$ & $-5.6^{+1.2}_{-0.9}$ & $5.4^{+1.6}_{-1.7}$ & $414^{+42}_{-41}$ & $389^{+39}_{-39}$ \\

    O\, {\sc vii}\footnote{The normalization and corrected flux values for the forbidden line of O\, {\sc vii} are based on the 68\% confidence limit for a Gaussian line profile with a fixed centroid and a fixed width (of 7 m\AA).  The data are consistent with a non-detection of this line.} & $22.0977$ (f) & \nodata  & \nodata & \nodata & $<19.3$ & $<17.9$  \\

    N\, {\sc vii}& $24.7810$ & \nodata & \nodata & $7.7_{-3.2}^{+3.7}$
    & $92\pm{21}$ & $75^{+18}_{-16}$ \\

    \hline

  \end{tabular}

\label{tab:Gaussian_fits}
\end{minipage}
\end{table*}

\begin{figure}
\includegraphics[angle=0,width=90mm]{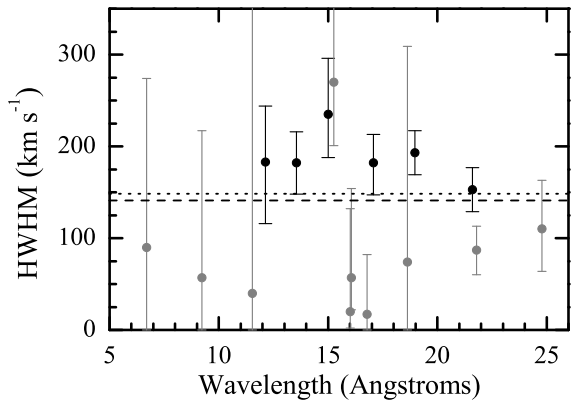}
\caption{Half-width at half-maximum line widths derived from fitting
  Gaussian line-profile models to each individual line in the MEG
  first-order spectrum. The strongest, well behaved, isolated lines
  are indicated with darker symbols.  The lower signal-to-noise,
  blended, or otherwise less reliable lines are shown in gray.  68\%
  confidence limit error bars are shown for all the lines.  The
  weighted mean HWHM velocity for all the lines is indicated by the
  dashed horizontal line, while the HWHM that corresponds to the
  turbulent velocity derived from the global spectral fit is indicated
  by the dotted horizontal line (recall that $v_{\mathrm {turb}} =
  180$ \kms\ corresponds to a half width at half maximum of 150 \kms).
}
\label{fig:linewidths}
\end{figure}

We first fit each emission line in the MEG spectrum individually,
fitting the negative and positive first order spectra simultaneously
(but not coadded) with a Gaussian profile model plus a power law to
represent the continuum.  In general, we fit the continuum near a line
separately to establish the continuum level, using a power-law index
of $n=2$, ($F_{\nu} \propto \nu^n$, which gives a flat spectrum in
$F_\lambda$ vs.\ $\lambda$).  We then fixed the continuum level and
fit the Gaussian model to the region of the spectrum containing the
line. We used the C statistic \citep{Cash1979}, which is appropriate
for data where at least some bins have very few counts, for all the
fitting of individual lines.  We report the results of these fits, and
the derived properties of the lines, in Tab.\ \ref{tab:Gaussian_fits},
and show the results for the line widths in Fig.\
\ref{fig:linewidths}.  Note that the half-width at half-maximum line
width, for the higher signal-to-noise lines (darker symbols in Fig.\
\ref{fig:linewidths}), is about 200 \kms. The weighted mean HWHM of
all the lines is 143 \kms, which is indicated in the figure by the
dashed line.  The typical thermal width for these lines is expected to
be roughly 50 \kms, so thermal broadening cannot account for the
observed line widths.



To explicitly test for the presence of broadening beyond thermal
broadening, when we fit the two-temperature variable abundance APEC
models to the entire MEG spectrum we accounted explicitly for
turbulent and thermal broadening.  We found a turbulent velocity of
$180$ \kms, corresponding\footnote{As implemented in ISIS, the
  Gaussian line profile half-width is given by $0.83\sqrt{v_{\mathrm
      {th}}^2 + v_{\mathrm {turb}}^2}$, so the HWHM that corresponds
  to the best-fit turbulent velocity value is 150 \kms.} to a
half-width at half-maximum turbulent velocity of 150 \kms, from a
single global fit to the spectrum.  We will characterize the overall
line widths as $\vhwhm=150$ \kms\ for the rest of the paper. The
thermal velocity for each line follows from the fitted temperature(s)
and the atomic mass of the parent ion.  This turbulent broadening is
consistent with what we find from the individual line fitting, as
shown in Fig.\ \ref{fig:linewidths}. We also find a shift of $-20$
\kms\/ from a single global fit to the spectrum. We note that this 20
\kms\/ blue shift of the line centroids is not significant, given the
uncertainty in the absolute wavelength calibration of the \hetgs\/
\citep{mdi2004}.

Because the emission lines are only barely resolved, there is not a
lot of additional information that can be extracted directly from
their profile shapes.  However, because the working assumption is that
this line emission arises in the stellar wind, it makes sense to fit
line profile models that are specific to stellar wind X-ray emission.
We thus fit the empirical wind-profile X-ray emission line model
developed by \citet{oc2001} to each line, again with a power-law
continuum model included.  This wind-profile model, though informed by
numerical simulations of line-driving instability (LDI) wind shocks
\citep{OCR1988,Cohen1996,Feldmeier1997}, is phenomenological, and only
depends on the spatial distribution of X-ray emitting plasma, its
assumed kinematics (described by a beta velocity law in a spherically
symmetric wind), and the optical depth of the bulk cool wind.  For the
low-density wind of \bcru, it is safe to assume that the wind is
optically thin to X-rays.  And once a velocity law is specified, the
only free parameter of this wind-profile model - aside from the
normalization - is the minimum radius, \Rmin, below which there is
assumed to be no X-ray emission.  Above \Rmin, the X-ray emission is
assumed to scale with the square of the mean wind density.

When we fit these wind-profile models, we had to decide what velocity
law to use for the model.  From CAK \citep{cak1975} theory, we expect
a terminal velocity of roughly 2000 \kms. The $\beta$ parameter of the
standard wind velocity law governs the wind acceleration according to
$v=\vinf(1-\Rstar/r)^{\beta}$ so that large values of $\beta$ give
more gradual accelerations.  Typically $\beta \approx 1$.  When we fit
the wind-profile model with $\vinf=2000$ \kms\ and $\beta=1$ to the
stronger lines, we could only fit the data if $\Rmin < 1.1$ \Rstar.
This puts nearly all of the emission at the slow-moving base of the
wind (because of the density-squared weighting of X-ray emissivity and
the $v^{-1}r^{-2}$ dependence of density), which is unrealistic for
any type of wind-shock scenario, but which is the only way to produce
the relatively narrow profiles in a model with a large wind terminal
velocity. This quantitative result verifies the qualitative impression
that the relatively narrow emission lines in the MEG spectrum are
inconsistent with a standard wind-shock scenario in the context of a
fast wind.


\begin{table*}
\begin{minipage}{180mm}
\caption{Helium-like complexes}
\begin{tabular}{ccccccc}
  \hline
  Ion & ${\mathcal R} \equiv f/i$\footnote{From Gaussian fitting. The values we obtain from the APEC global fitting within ISIS are consistent with the $f/i$ values reported here.} & $\lambda_{\rm UV}$ (\AA)\footnote{The UV wavelengths at which photoexcitation out of the upper level of the forbidden line occurs.}  & $H_{\nu}$ (ergs s$^{-1}$ cm$^{-2}$ Hz$^{-1}$)\footnote{The assumed photospheric Eddington flux at the relevant UV wavelengths, based on the TLUSTY model atmosphere, which we use to calculate the dependence of $f/i$ on the radius of formation, \Rfir.}  & \Rfir~(\Rstar)\footnote{Formation radius using eqn.\ (1c) of \citet{bdt1972} and eqn.\ (3) of \citet{Leutenegger2006}, and based on the $f/i$ values in the second column.}  & \Rmin~(\Rstar)\footnote{From wind profile fitting, assuming $\beta=1$ and letting \vinf\/ be a free parameter. The ${\mathcal G}$ values in the last column also come from the profile fitting. } & ${\mathcal G} \equiv \frac{(f+i)}{r}$  \\
  \hline
  Ne\, {\sc ix} & $0.09 \pm .05$ & $1248, 1273$ & $1.35 \times 10^{-3}$ & $2.9^{+0.7}_{-1.0}$  & $1.37^{+.18}_{-.02}$ & 1.05 \\
  Mg\, {\sc xi}  & $0.65 \pm .44$  & $998, 1034$ & $1.39 \times 10^{-3}$ & $3.2^{+1.4}_{-1.5}$  & $1.80^{+1.33}_{-0.45}$ & 1.01 \\

\hline
\end{tabular}
\label{tab:fir}
\end{minipage}
\end{table*}


\begin{figure}
\includegraphics[angle=90,width=80mm]{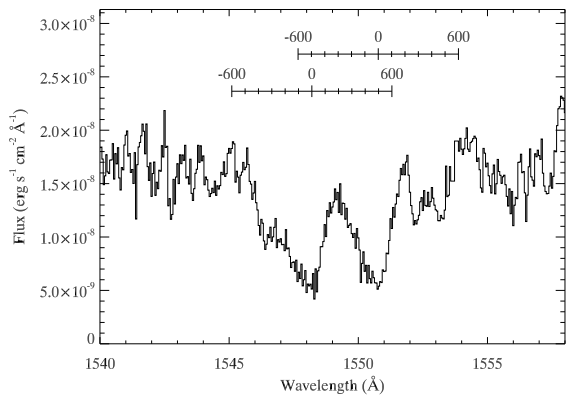}
\caption{The C\,{\sc iv} resonance doublet from several coadded
  archived \iue\/ observations.  A velocity scale (km s$^{-1}$) in the
  frame of the star is shown above the data.  These lines show the
  characteristic blue-shifted absorption that is expected from a
  stellar wind, but the lines are not only weak compared to
  theoretical expectations but also much less wind broadened.  }
\label{fig:CIV}
\end{figure}


\begin{figure}
\includegraphics[angle=90,width=80mm]{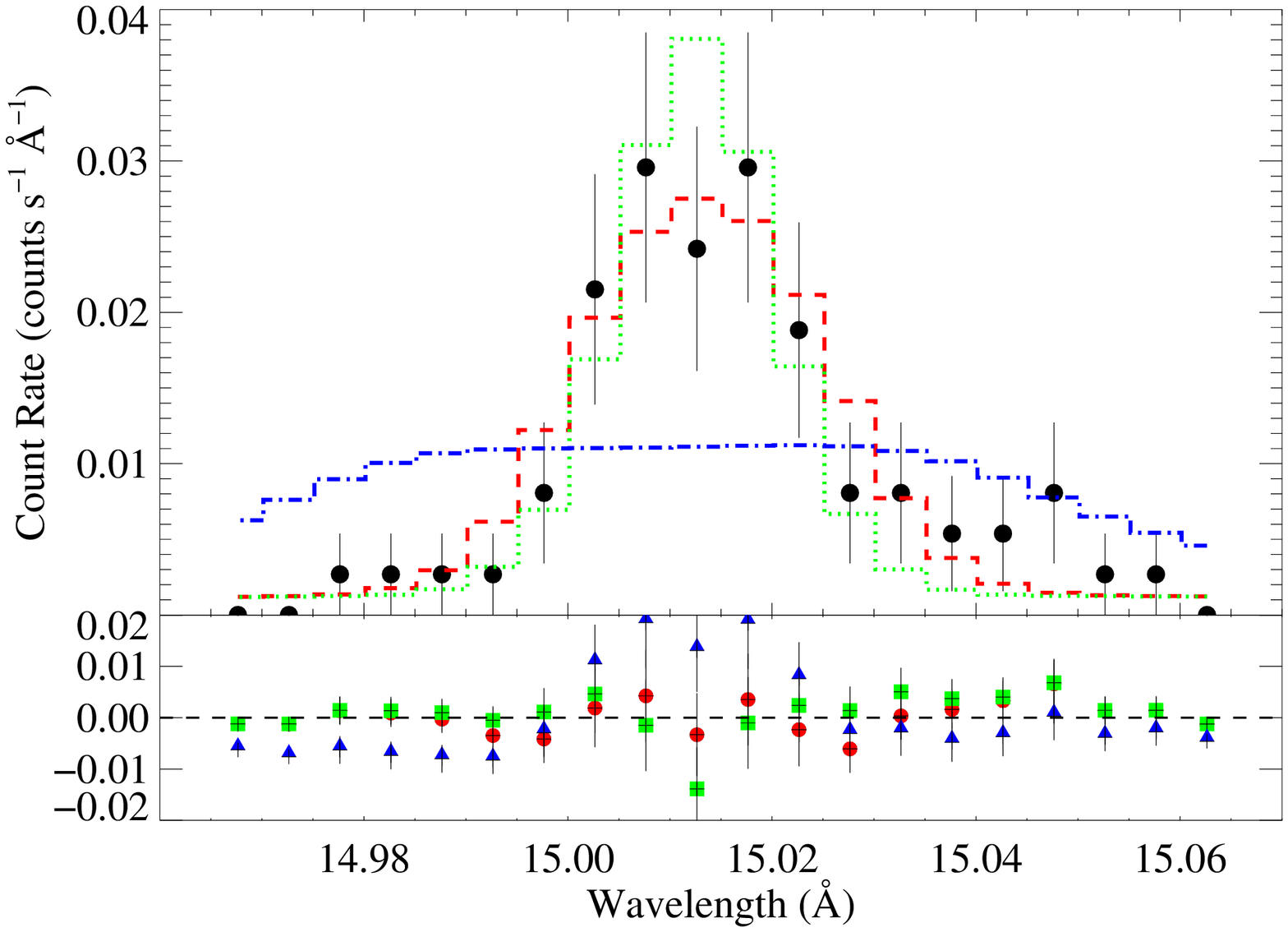}
\caption{The Fe\, {\sc xvii} line at 15.014 \AA\ with three different
  wind profile models. The red, dashed line is a constant-velocity
  wind model with $\vinf = 280$ \kms\ (the best-fit value for a
  constant outflow velocity, which produces emission lines with
  $\vhwhm \approx 150$ \kms). The model with $\vinf=2000$ \kms,
  $\beta=1$, and $\Rmin=1.5$ \Rstar\ is represented by the blue,
  dot-dash line.  An infinitely narrow model is shown as the green,
  dotted line. The residuals for each model fit are shown in the lower
  panel, as red circles for the global best-fit, modestly broadened
  ($\vinf = 280$ \kms) model, green squares for the narrow profile
  model, and blue triangles for the broad wind model.  The wind model
  with the higher velocity ($\vinf=2000$ \kms) clearly does not
  provide a good fit, while the narrower constant velocity ($\vinf =
  280$ \kms) wind model does.  And while the very narrow model cannot
  be absolutely ruled out, the $\vinf = 280$ \kms\/ model is preferred
  over it with a high degree of significance. All the models shown
  here have been convolved with the instrumental response function
  ({\it rmf}) and multiplied by the wavelength-dependent effective
  area ({\it garf}).  }
\label{fig:windprof}
\end{figure}

However, although the theoretical expectation is for a fast wind with
a terminal velocity of roughly 2000 \kms, the UV wind line profiles
observed with \iue\/ tell a different story.  The UV line with the
strongest wind signature is the C\,{\sc iv} doublet at 1548,1551 \AA,
which shows a blue edge velocity of only 420 \kms\ \citep{Prinja1989}.
We show this line in Fig.\ \ref{fig:CIV}. The Si\, {\sc iii} line
shows no wind broadening and the Si\, {\sc iv} line shows a slightly
weaker wind signature than the C\, {\sc iv} feature.  Certainly the
terminal velocity of the wind could be significantly higher than this,
with the outer wind density being too low to cause noticeable
absorption at high velocities. However, because the empirical evidence
from these UV wind features indicates that the terminal velocity of
the wind may actually be quite low, we refit the strong lines in the
MEG spectrum with the wind-profile model, but this time using a
terminal velocity of 420 \kms.  The fits were again statistically
good, and for these models, the fitted values of \Rmin\ were generally
between 1.3 \Rstar\ and 1.5 \Rstar, which are much more reasonable
values for the onset radius of wind-shock emission.


We also considered a constant-velocity model ($\beta=0$), where the
X-rays perhaps are produced in some sort of termination shock.  For
these fits, \Rmin\/ hardly affects the line width, so we fixed it at
1.5 \Rstar\ and let the (constant) terminal velocity, \vinf, be a free
parameter.  For the seven lines or line complexes with high enough
signal-to-noise to make this fitting meaningful, we found best-fit
velocities between 270 \kms\ and 300 \kms\ for five of them, with one
having a best-fit value lower than this range and one line having a
best-fit value higher than this range. The average wind velocity for
these seven lines is $\vinf = 280$ \kms.  All in all, the emission
line widths are consistent\footnote{The measured line widths (of
  $\vhwhm \approx 150$ \kms\/ derived from both the Gaussian fitting
  and the global thermal spectral modeling) are expected to be
  somewhat smaller than the modeled wind velocity, as some of the
  wind, in a spherically symmetric outflow, will be moving
  tangentially to the observer's line of sight. So, a spherically
  symmetric, constant velocity wind with $\vinf=280$ \kms\/ is
  completely consistent with emission lines having HWHM values of 150
  \kms.} with a constant-velocity outflow of $280$ \kms, and are
inconsistent with a wind-shock origin in a wind with a terminal
velocity of 2000 \kms.  The lines are narrower than would be expected
even for a wind with a terminal velocity not much greater than the
observed C\, {\sc iv} blue edge velocity, $v = 420$ \kms\ (based on
our $\beta=1$ fits as well as the $\beta=0$ fits). The X-ray emitting
plasma must be moving significantly slower than the bulk wind. In
Fig.\ \ref{fig:windprof} we show the best-fit constant velocity model
($\vinf = 280$ \kms) superimposed on the Fe\, {\sc xvii} line at
15.014 \AA, along with the higher velocity model based on the
theoretically expected terminal velocity, which clearly does not fit
the data well. We also compare the best-fit wind-profile model to a
completely narrow profile in this figure.  The modestly wind-broadened
profile that provides the best fit is preferred over the narrow
profile at the 99\% confidence level, based on the C statistic values
of the respective fits ({\it Numerical Recipes}, Ch.\ 14;
\citet{Press1982}).


\subsubsection{Helium-like forbidden-to-intercombination line ratios}

The final spectral diagnostic we employ is the UV-sensitive
forbidden-to-intercombination line ratio of helium-like ions.  These
two transitions to the ground-state have vastly different lifetimes,
so photoexcitation out of the $1{\rm s}2{\rm s}~^3{\rm S}_{1}$ state,
which is the upper level of the forbidden line, to the $1{\rm s}2{\rm
  p}~^3{\rm P}_{1,2}$ state, which is the upper level of the
intercombination line, can decrease the forbidden-to-intercombination
line ratio, ${\mathcal R} \equiv f/i$.  Because the photoexcitation
rate depends on the local UV mean intensity, the $f/i$ ratio is a
diagnostic of the distance of the X-ray emitting plasma from the
photosphere. This diagnostic has been applied to many of the hot stars
that have \chandra\/ grating spectra, and typically shows a source
location of at least half a stellar radius above the photosphere.
Recently, \citet{Leutenegger2006} have shown that the $f/i$ ratios
from a spatially distributed X-ray emitting plasma can be accounted
for self-consistently with the line profile shapes in four O stars,
providing a unified picture of wind-shock X-ray emission on O stars.





\begin{figure}
\includegraphics[angle=0,scale=0.35,width=80mm]{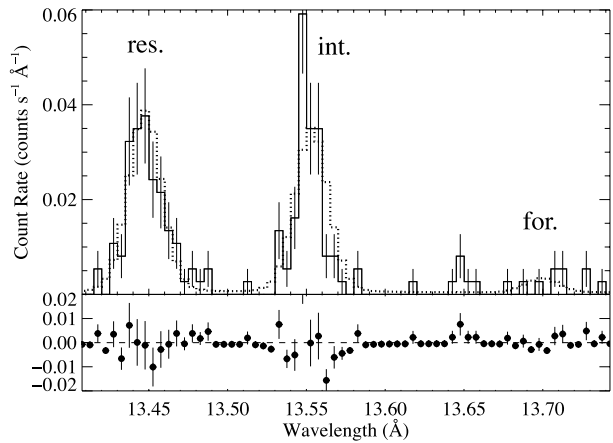}
\includegraphics[angle=0,width=85mm]{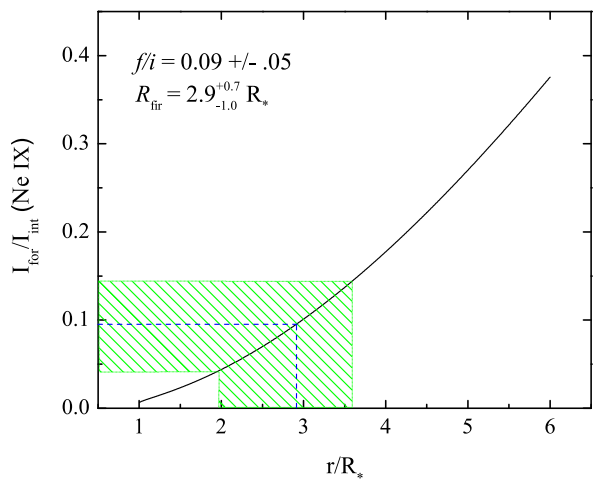}
\caption{The Ne\, {\sc ix} resonance-intercombination-forbidden
  complex in the co-added MEG data, with the best-fit three-Gaussian
  model overplotted (top). The constraints on the measured $f/i$ ratio
  ($0.09 \pm .05$), from the fit shown in the top panel, are indicated
  by the cross-hatched area's intersection with the y-axis (bottom).
  The solid black curve is the model for the line ratio, as a function
  of radius.  The cross-hatched region's intersection with the model
  defines the 68\% confidence limit on $r/\Rstar$ of (1.9:3.6).  The
  dashed line represents the best-fit value, $f/i = 0.09$, and the
  corresponding $\Rfir=2.9$ \Rstar.  }
  \label{fig:fir_neix}
\end{figure}


\begin{figure}
\includegraphics[angle=0,scale=0.35,width=80mm]{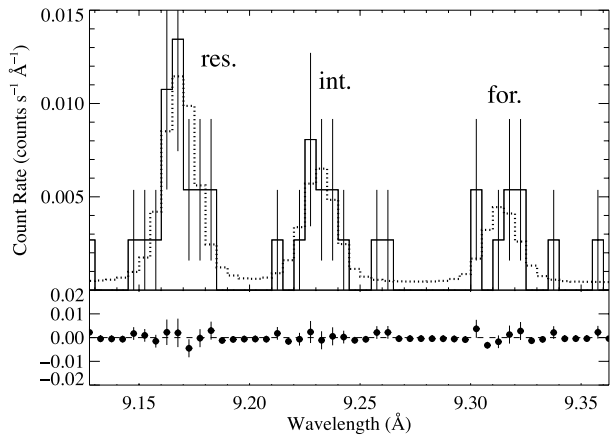}
\includegraphics[angle=0,width=85mm]{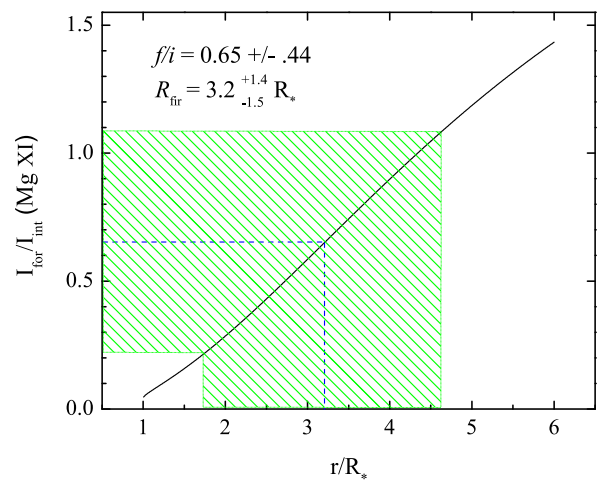}
\caption{The Mg\, {\sc xi} resonance-intercombination-forbidden
  complex in the co-added MEG data, with the best-fit wind-profile
  models overplotted (top).  The constraints on the formation radius,
  \Rfir, are shown in the bottom panel.  }
\label{fig:fir_mgxi}
\end{figure}

The lower-Z species like oxygen do not provide any meaningful
constraints, as is the case for O stars, and there are not enough
counts in the Si\, {\sc xiii} complex to put any meaningful
constraints on the source location.  Only the Ne\, {\sc ix} and Mg\,
{\sc xi} complexes provide useful constraints.  For both Ne\, {\sc ix}
and Mg\, {\sc xi} we find $\Rfir \approx 3 \pm 1 \Rstar$. Of course,
the line-emitting plasma is actually distributed in the wind, and so
we also fit the same modified \citet{oc2001} wind profile model that
\citet{Leutenegger2006} used to fit a distributed X-ray source model
to the He-like complexes, accounting simultaneously for line ratio
variations and wind-broadened line profiles. With the wind velocity
law and ${\mathcal G} \equiv \frac{f + i}{r}$ as free parameters,
strong constraints could not be put on the models. Typically, these
distributed models of \ftoi\/ gave onset radii, \Rmin, that were quite
a bit lower than \Rfir.  This is not surprising, because a distributed
model will inevitably have some emission closer to the star than
\Rfir\ to compensate for some emission arising at larger radii.  We
list the results of such a wind-profile fit to the He-like complexes
in Tab.\ \ref{tab:fir} along with the \Rfir\/ values. The two
helium-like complexes and the best-fit models are shown in Figs.\
\ref{fig:fir_neix} and \ref{fig:fir_mgxi}, along with the modeling of
the $f/i$ ratios and the characteristic formation radii, \Rfir. We
note that the data rule out plasma right at the photosphere only at
the 68\% confidence level.  

We derive a single characteristic radius of formation, \Rfir, for each
complex by calculating the dependence of the $f/i$ ratio on the
distance from the photosphere, using a TLUSTY model atmosphere
\citep{LH2007}.  We used the formalism of \citet{bdt1972}, equation
(1c), ignoring collisional excitation out of the $^3{\rm S}_1$ level,
and explicitly expressing the dependence on the dilution factor, and
thus the radius, as in equation (3) in \citet{Leutenegger2006}. The
measurement of $f/i$ from the data can then be used to constrain the
characteristic formation radius, \Rfir.

\subsection{Time Variability Analysis}

We first examined the \chandra\ data for overall, stochastic
variability.  Both a K-S test applied to the unbinned photon arrival
times and a ${\chi}^2$ fitting of a constant source model to the
binned light curve showed no evidence for variability in the combined
zeroth order and first order spectra.  This is to be expected from
wind-shock X-rays, which generally show no significant variability.
This is usually interpreted as evidence that the X-ray emitting plasma
is distributed over numerous spatial regions in the stellar wind
\citep{CCM1997}.  Separate tests of the hard ($h\nu>1$ keV) and soft
($h\nu<1$ keV) counts showed no evidence for variability in the soft
data, but some evidence for variability (K-S statistic implies 98\%
significance) in the hard data.

Because \bcru\/ is a \bcep\/ variable, we also tested the data for
periodic variability.  Both the changes in the star's effective
temperature with phase and the possible effects of wave leakage of the
pulsations into the stellar wind could cause the X-ray emission, if it
arises in wind shocks, to show a dependence on the pulsation phase
\citep{OC2002}.  Our \chandra\ observation covers about four pulsation
periods.  To test for periodic variability, we used a variant of the
K-S test, the Kuiper test \citep{Paltani2004}, which is also applied
to unbinned photon arrival times. To test the significance of a given
periodicity, we converted the photon arrival times to phase and formed
a cumulative distribution of arrival phases from the events table.  We
then calculated the Kuiper statistic and its significance for that
particular period.  By repeating this process for a grid of test
periods, we identified significant periodicities in the data.


\begin{figure}
\includegraphics[angle=90,width=80mm]{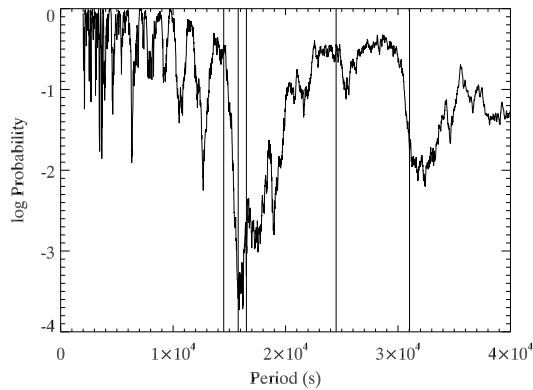}
\caption{Rejection probability (significance level increasing
  downward) of assumed period, according to the Kuiper statistic.  The
  five known optical pulsation periods \citep{Aerts1998,Cuypers2002}
  are indicated by the vertical lines. The primary pulsation period is
  the third of these.}
\label{fig:periodogram}
\end{figure}


\begin{figure}
\includegraphics[angle=90,width=80mm]{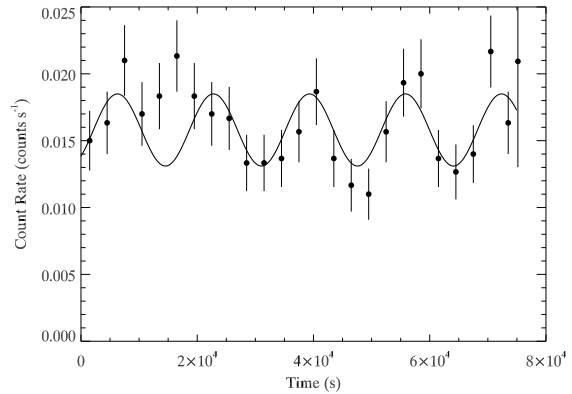}
\caption{The light curve of the hard counts in the zeroth and first order
  spectra of the primary, along with the best-fit sinusoidal model.
}
\label{fig:lightcurve}
\end{figure}

When we applied this procedure to all the zeroth-order and first-order
counts we did not find any significant periodicities.  However,
applying it only to the hard counts produced a significant peak near
the primary and tertiary optical pulsation periods ($f_1$ and $f_3$)
of 4.588 hours and 4.386 hours \citep{Aerts1998}, as shown in Fig.\
\ref{fig:periodogram}.  This peak is significant at the 99.95\% level.
It is relatively broad as the data set covers only about four cycles
of the pulsation, so it is consistent with both pulsation modes, but
not with the secondary ($f_2$) mode \citep{Aerts1998}. No significant
periods are found in the soft bandpass, despite the higher
signal-to-noise there.  To quantify the level of periodic variability
we found in the hard counts, we made a binned light curve from these
counts and fit a sine wave with a period of 4.588 hours, leaving only
the phase and the amplitude as free parameters.  We find a best-fit
amplitude of 18\%.  We show the light curve and this fit in Fig.\
\ref{fig:lightcurve} and note that there appears to be some additional
variability signal beyond the strictly periodic component. We also
note that there is a phase shift of about a quarter period
($\Delta\phi = 0.27 \pm .04$) between the times of maximum X-ray and
optical light, with the X-rays lagging behind the optical variability.
To make this assessment, we compared the time of maximum X-ray light
from the sinusoidal fitting shown in Fig.\ \ref{fig:lightcurve} to the
time of maximum optical light in the WIRE data, based on just the
primary pulsation mode (Cuypers 2007, private communication).  We note
that $\phi=0.25$ corresponds to the maximum blue shift in the radial
velocity variability of \bcep\ stars \citep{mgc1992}.



\section{Analysis of the companion} \label{sec:companion}

We subjected the newly discovered companion to most of the same
analyses we have applied to \bcru, as described in the previous
section. The exceptions are the wind-profile fitting, which is not
relevant for the unresolved lines in the companion's spectrum, and the
tests for periodic variability, since none is expected.


\begin{figure}
\includegraphics[angle=90,width=80mm]{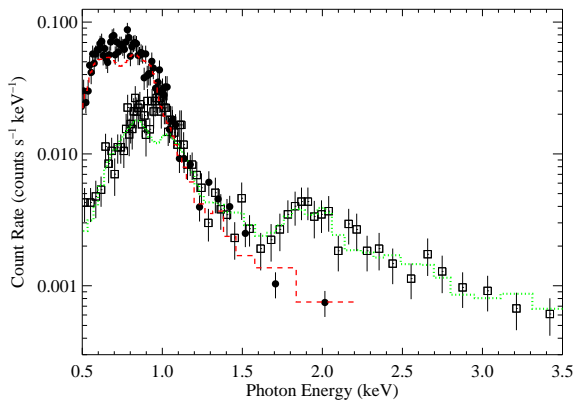}
\caption{The zeroth order spectrum of the companion (open squares) is
  shown along with that of \bcru\/ (filled circles).  The companion's
  spectrum is significantly harder, and is well-fit by a
  two-temperature thermal model with component temperatures several
  times higher than those found for \bcru. The best-fit models -- to
  the combined zeroth order and grating spectra -- are shown as the
  dashed (\bcru) and dotted (companion) lines.}
\label{fig:zeroth_order_both}
\end{figure}


\begin{figure*}
\includegraphics[angle=0,width=180mm]{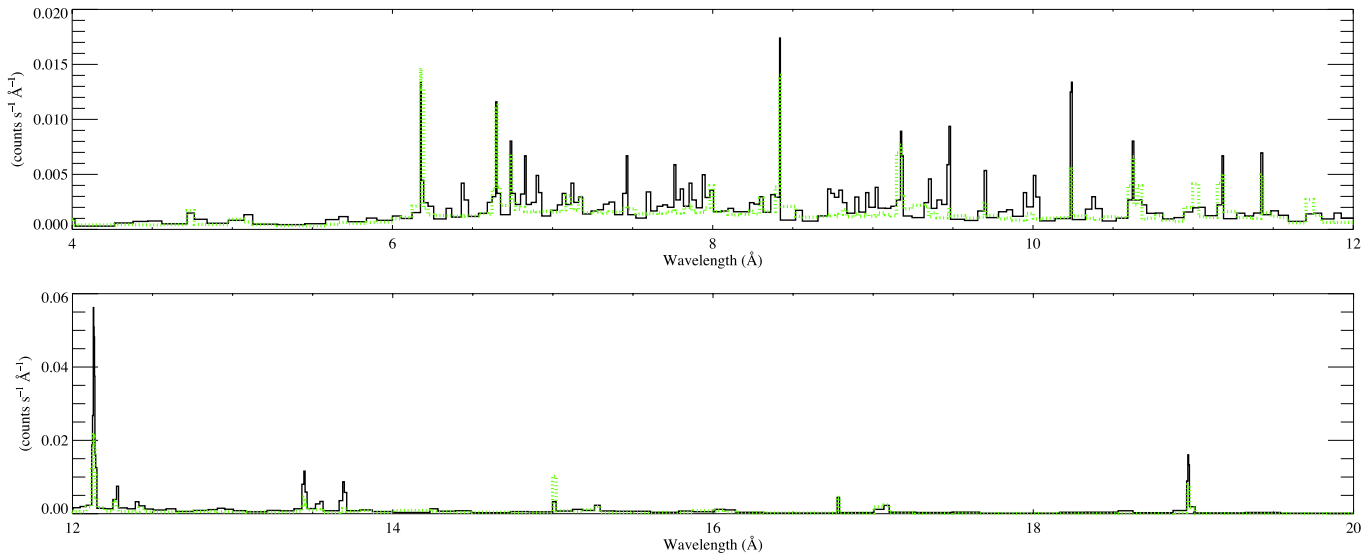}
\caption{The best-fit two-temperature, variable abundance thermal
  spectral (APEC) model (green, dotted line), superimposed on the
  adaptively smoothed co-added MEG first-order spectrum of the
  companion (black, solid line).  }
\label{fig:companion_MEG_apec}
\end{figure*}

Our working hypothesis is that the X-ray bright companion is a
low-mass pre-main-sequence star, similar to those found in the
\rosat\/ pointing \citep{pf1996,fl1997,Alcala2002}.  In the following
two subsections, we marshal evidence from the spectral and
time-variability properties to address this hypothesis.

\subsection{Spectral Analysis}

The spectrum of the companion is significantly harder than that of
\bcru.  This can be seen in both the zeroth-order spectrum (see Fig.\
\ref{fig:zeroth_order_both} and Fig.\ \ref{fig:finding_chart}) and the
dispersed spectrum (see Fig.\ \ref{fig:companion_MEG_apec}).  The
dispersed spectrum has quite poor signal-to-noise, as the hardness
leads to fewer counts per unit energy flux.



We fit a two-temperature optically thin coronal equilibrium spectral
model to the zeroth order spectrum of the companion. As we did for the
primary, we compared the best-fit model's predictions to the MEG
spectrum and found good agreement. This source is significantly harder
than \bcru, as indicated by the model fit, which has component
temperatures of 6.2 and 24 million K. The emission measures of the two
components are $1.8 \times 10^{52}$ cm$^{-3}$ and $4.2 \times 10^{52}$
cm$^{-3}$, respectively, yielding an X-ray flux of $5.71 \pm .06
\times 10^{-13}$ ergs s$^{-1}$ cm$^{-2}$, corrected for ISM absorption
using the same column density that we used for the analysis of the
spectrum of \bcru.  This flux implies $\lx = 8.0 \times 10^{29}$ ergs
s$^{-1}$.  No significant constraints can be put on the abundances of
the X-ray emitting plasma from fitting the \chandra\/ data. This
best-fit two-temperature APEC model is shown superimposed on the
adaptively smoothed, coadded MEG spectrum in Fig.\
\ref{fig:companion_MEG_apec} (it is also shown, as the dotted
histogram, in Fig.\ \ref{fig:zeroth_order_both}). These MEG data are
much noisier than those from \bcru, but the overall fit is reasonable,
and the stronger continuum, indicative of higher temperature thermal
emission is quite well fit. We note that only three of the line
complexes have even a dozen counts in this spectrum.

Unlike \bcru, the companion has lines that are narrow -- unresolved at
the resolution limit of the \hetgs. This, and the hardness of the
spectrum, indicates that the companion is consistent with being an
active low-mass main-sequence or pre-main-sequence star.

The $f/i$ ratios that provided information about the UV mean intensity
and thus the source distance from the photosphere in the case of
\bcru\/ can be used in coronal sources as a density diagnostic.  The
depopulation of the upper level of the forbidden line will be driven
by collisions, rather than photoexcitation, in the case of cooler
stars. Some classical T Tauri stars (cTTSs) have shown anomalously low
$f/i$ ratios compared to main sequence stars and even weak-line T
Tauri stars (wTTSs) \citep{Kastner2002,Schmitt2005}. The best
constraint in the companion's spectrum is provided by Ne\, {\sc ix},
which has $f/i = 2.2 \pm 1.1$.  This value is consistent with the
low-density limit of $f/i = 3.0$ \citep{pd2000}, which is typically
seen in wTTSs and active main sequence stars but is sometimes altered
in cTTSs.  Using the calculations in \citet{pd2000}, we can place an
upper limit on the electron density of $n_{\rm e} \approx 10^{12}$
cm$^{-3}$, based on the $1 \sigma$ lower limit of the $f/i$ ratio.

\subsection{Time Variability Analysis}


\begin{figure}
\includegraphics[angle=90,width=80mm]{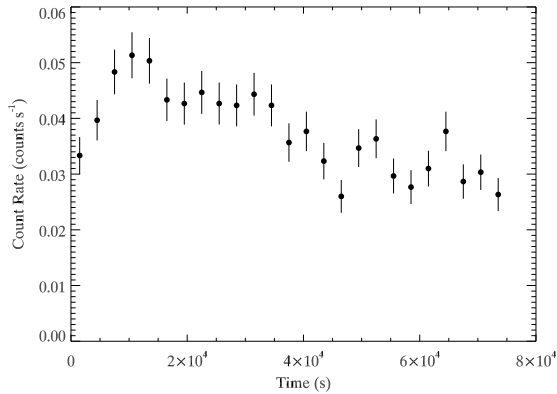}
\caption{The binned light curve of the companion, including all
  zeroth- and first-order counts. }
\label{fig:companion_lightcurve}
\end{figure}


\begin{figure}
\includegraphics[angle=90,width=80mm]{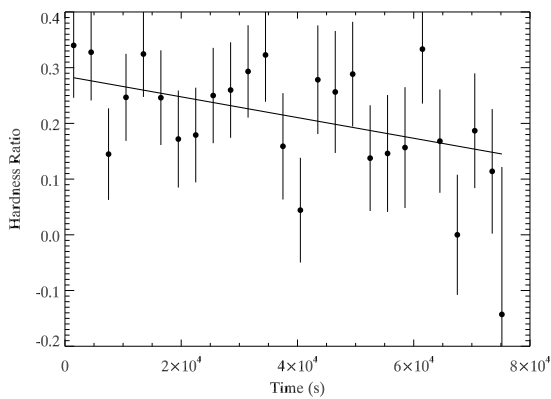}
\caption{The hardness ratio as a function of time, defined as $HR
  \equiv \frac{CR{\mathrm {(>1keV)}} - CR{\mathrm
      {(<1keV)}}}{CR{\mathrm {(>1keV)}} + CR{\mathrm {(<1keV)}}}$,
  where $CR$ denotes the count rate in the combined zeroth order and
  first order (MEG) data, over the indicated photon energy ranges.  We
  fit a linear function to the data, and find $\chi^2=25$ for 24
  degrees of freedom for a constant model (slope of zero), so a
  constant source model cannot be rejected outright.  However, the
  best-fit slope of $-1.9 \times 10^{-6}$ cts s$^{-2}$ (solid line)
  gives $\chi^2=20$, for 23 degrees of freedom, which is a
  statistically significant improvement over the constant model.}
\label{fig:companion_hardness}
\end{figure}

The companion is clearly much more variable than is \bcru.  The K-S
test of the unbinned photon arrival times compared to a constant
source model shows significant variability in all energy bands, though
the significance is higher in the hard X-rays ($h\nu > 1$ keV). The
binned light curve also shows highly significant variability, again
with the null hypothesis of a constant source rejected for the overall
data, and both the soft and hard counts, separately. An overall
decrease in the count rate is clearly seen as the observation
progressed (Fig.\ \ref{fig:companion_lightcurve}).



The short rise followed by a steady decline of nearly 50\% in the
X-ray count rate over the remainder of the observation is
qualitatively consistent with flare activity seen in late-type PMS
stars \citep{Favata2005}, although the rise is not as rapid as would
be expected for a single, large flare.  We examined the spectral
evolution of the companion during this decay, and there is mild
evidence for a softening of the spectrum at late times, but not the
dramatic spectral evolution expected from a cooling magnetic loop
after a flare. For example, the hardness ratio (the difference between
the count rate above 1 keV and below 1 keV, normalized by the total
count rate) is shown in Fig.\ \ref{fig:companion_hardness}.  There is
a mild preference for a negative slope over a constant hardness ratio
(significant at the 95\% level), but with the data as noisy as they
are, a constant hardness ratio cannot be definitively ruled out.

\section{Discussion and Conclusions} \label{sec:discussion}

Although it is very hot and luminous, \bcru\/ has a \chandra\/
spectrum that looks qualitatively quite different from those of O
stars, with their significantly Doppler broadened X-ray emission
lines.  Specifically, the emission lines of \bcru\/ are quite narrow,
although they are resolved in the MEG, with half-widths of $\sim 150$
\kms.  These relatively narrow lines are incompatible with the
wind-shock scenario that applies to O stars if the wind of \bcru\/ has
a terminal velocity close to the expected value of $\vinf \approx
2000$ \kms. They are, however, consistent with a spherically symmetric
outflow with a constant velocity of $\vinf=280$ \kms.

If the wind terminal velocity is significantly less than 2000 \kms\/
-- perhaps just a few hundred \kms\/ above the $420$ \kms\/ maximum
absorption velocity shown by the strongest of the observed UV wind
lines -- a wind-shock origin of the X-ray emission is possible, but
still difficult to explain physically. One wind-shock scenario that we
have considered relies on the fact that the low-density wind of
\bcru\/ should take a long time to cool once it is shock heated. If a
large fraction of the accelerating wind flow passes through a
relatively strong shock front at $r \approx 1.5$ \Rstar, where the
pre-shock wind velocity may be somewhat less than 1000 \kms, it could
be decelerated to just a few 100 \kms, explaining the rather narrow
X-ray emission lines. Similar onset radii are seen observationally in
O stars \citep{kco2003,Cohen2006,Leutenegger2006}.  The hottest plasma
component from the global thermal spectral modeling requires a shock
jump velocity of roughly 700 \kms, with the dominant cooler component
requiring a shock jump velocity of roughly 400 \kms. 

The \ftoi\/ ratios of Ne\, {\sc ix} and Mg\, {\sc xi} indicate that
the hot plasma is either several stellar radii from the photosphere or
distributed throughout the wind starting at several tenths of a
stellar radius.  This is broadly consistent with this scenario of an
X-ray emitting outer wind. The lack of overall X-ray variability would
also be broadly consistent with this quasi-steady-state wind-shock
scenario.  And the periodic variability seen in the hard X-rays could
indicate that the shock front conditions are responding to the stellar
pulsations, while the lack of soft X-ray variability would be
explained if the softer X-rays come from cooling post-shock plasma in
which the variability signal has been washed out.

The overall level of X-ray emission is quite high given the modest
wind density, which is a problem that has been long-recognized in
early B stars \citep{CCM1997}.  A large fraction of the wind is
required to explain the X-ray emission.  Elaborating on the scenario
of a shock-heated and inefficiently cooled outer wind, we can crudely
relate the observed X-ray emission measure to the uncertain mass-loss
rate.

Assuming a spherically symmetric wind in which the X-rays arise in a
constant-velocity post-shock region that begins at a location, $R_0$,
somewhat above the photosphere; say $R_0 \approx 1.5$ (in units of
\Rstar), we can calculate the emission measure available for X-ray
production from

\[
\EM \equiv \int n_{\rm e} n_{\rm H} dV = \int^{\infty}_{R_0} n_{\rm e} n_{\rm H} 4 \pi r^2 dr. 
\]

\noindent
Again assuming spherical symmetry and a smooth flow with a constant
velocity for $r > R_0$, we have

\[
\EM = 1.3 \times 10^{54}\frac{\Mdotnine^2}{R_0\vhundred^2} ~~~ {\rm (cm^{-3})}
\]

\noindent
for the emission measure of the entire wind above $r=R_0$, where
\Mdotnine\/ is the mass-loss rate in units of $10^{-9} \Msunyr$ and
\vhundred\/ is the wind velocity in units of $100$ \kms. Equating this
estimate of the available emission measure to that which is actually
observed in the data, $3.5 \times 10^{53}$ cm$^{-3}$, we get

\[
\Mdotnine=\sqrt{\frac{\EM}{1.3 \times 10^{54}}R_0\vhundred^2}
\approx 1
\] 

\noindent
for $R_0 \approx 1.5$ and $v = 280$ \kms\ ($\vhundred = 2.8$), which
is the velocity obtained from fitting the line widths in the MEG
spectrum.  Thus a mass-loss rate of order $10^{-9} \Msunyr$ is
consistent with the observed X-ray emission measure\footnote{We hasten
  to point out that our analysis ignores the possibility that less
  than the entire wind volume beyond $r = R_0$ is X-ray emitting.
  Certainly a filling factor term could multiply the mass-loss rate in
  the expression for the emission measure.  Ignoring such a finite
  filling factor would lead us to underestimate the mass-loss rate.
  However, another simplification we have made here involves the
  assumption of a smooth, spherically symmetric distribution.  Any
  clumping, or deviation from smoothness, in the post-shock region
  will lead to more emission measure for a given mass-loss rate (and
  assumed $R_0$ and \vhundred).  Ignoring this effect will lead us to
  overestimate the mass-loss rate.  Thus the mass-loss rate we
  calculate here of $\Mdot \approx 10^{-9}$ \Msunyr\/ should be taken
  to be a crude estimate, subject to a fair amount of uncertainty.
  However, the two major oversimplifications in this calculation will
  tend to cancel each other out, as we have just described.  We thus
  conclude that a mass-loss rate of order $10^{-9}$ \Msunyr\/ is
  sufficient to explain the observed X-ray emission, assuming a
  constant, slow outflow of the post-shock plasma above an onset
  radius of about 1.5 \Rstar.  This mass-loss rate is lower than the
  predictions of CAK theory but significantly larger than the observed
  mass loss in the C\,{\sc iv} UV resonance line, implying that the
  actual mass-loss rate of \bcru\/ is in fact lower than CAK theory
  predicts, but that some of the weakness of the observed UV wind
  lines is due to ionization effects.  Of course, the presence of a
  large X-ray emitting volume in the outer wind provides a ready
  source of ionizing photons that can easily penetrate back into the
  inner, cool wind and boost the ionization state of metals in the
  wind.}.  This type of simple analysis relating
the observed X-ray emission measure in early B stars to their
mass-loss rate in the context of shock heating and inefficient
radiative cooling in the outer wind was applied to \rosat\/
observations of the early B giants $\epsilon$ and $\beta$ CMa by
\citet{ddh1994}, who, like we have here, found lower mass-loss rates
than expected from theory.  A downward revision in the mass-loss rates
of early B stars might not be surprising given recent work indicating
that the mass-loss rates of O stars have been overestimated
\citep{Bouret2005,fmp2006,Cohen2007}.

Given the estimated mass-loss rate of $\Mdot \approx 10^{-9}$
\Msunyr\/ for \bcru, we can assess the cooling time, or cooling
length, of the purported wind shocks.  A crude estimate of the cooling
time and length of shock-heated wind material can be made by comparing
the thermal energy content of the post-shock plasma, 

\[
E_{\mathrm {th}} = \frac{3}{2}n{\rm
  k}T \approx 0.1  ~~~~{\rm (ergs~cm^{-3})} 
\]

\noindent
to the radiative cooling rate,

\[
\frac{dE}{dt} = n^2\Lambda \approx 5 \times 10^{-7} ~~~~{\rm (ergs~s^{-1}~cm^{-3})}.
\]

\noindent
These numbers come from taking expected values of the wind density
based on the assumptions of spherical symmetry of the wind and a
constant outflow velocity, 

\[
n = 6 \times 10^8 \frac{\Mdotnine}{R^2\vhundred} ~~~~{\rm (cm^{-3}),}
\]

\noindent
where $R$ is the radial location in units of stellar radii.  Using
$\vhundred = 2.8$ and $\Mdotnine = 1$, the density is just $n \approx
10^8$ cm$^{-3}$ at $R = 1.5$. We use temperatures of $T = 2.5$ and
$6.5 \times 10^6$ K, given by the spectral fitting discussed in \S4.1,
assuming that some of the softer thermal component is produced
directly by shock heating (rather than by radiative cooling of hotter
plasma) since the emission measure of the hot component is so small.
We take the integrated line emissivity to be $\Lambda \approx 5 \times
10^{-23}$ ergs s$^{-1}$ cm$^3$ for a plasma of several million K
\citep{Benjamin2001}. The characteristic cooling time thus derived,

\[
t_{\mathrm {cool}} = \frac{3{\rm k}T}{2n\Lambda}
\approx 1 ~{\mathrm {to}}~ 2 \times 10^5 ~~~~{\rm (s),}
\]

\noindent
is many times longer than the characteristic flow time,

\[
t_{\mathrm {flow}} \equiv \frac{r}{v} \approx
\frac{\Rstar}{v} \approx 2 \times 10^4 ~~~~ {\rm (s).}
\]

\noindent
Thus, material shock heated to several million degrees at half a
stellar radius above the surface will essentially never cool back down
to the ambient wind temperature by radiative cooling, motivating the
picture we have presented of a wind with an inner, cool acceleration
zone and a quasi-steady-state outer region that is hot, ionized, X-ray
emitting, and moving at a more-or-less constant velocity (because it
is too highly ionized to be effectively radiatively driven).

Although the scenario we have outlined above is phenomenologically
plausible, there are several significant problems with it.  First of
all, the behavior of wind shocks produced by the line-driving
instability is not generally as we have described here.  Such shocks
tend to propagate outward at an appreciable fraction of the ambient
wind velocity \citep{OCR1988,Feldmeier1997}.  Whereas the scenario we
have outlined involves a strong shock that is nearly stationary in the
frame of the star, as one would get from running the wind flow into a
wall.  There is, however, no wall for the wind to run into. It is
conceivable that a given shock front forms near $r = 1.5$ \Rstar\/ and
propagates outward, with another shock forming behind it (upstream) at
roughly the same radius, so that the ensemble of shocks is close to
steady state.  However, the narrow observed X-ray lines severely limit
the velocity of the shock front, as the post-shock velocity in the
star's frame must be greater the shock front's velocity.  Yet the
observed lines are consistent with a post-shock flow of only $v = 280$
\kms. A related, second, problem with the scenario is that the shocks
required to heat the plasma to $T \approx 2.5$ to $6.5 \times 10^6$ K
are relatively strong for a weak wind.  Smaller velocity dispersions
are seen in statistical analyses of the LDI-induced structure in O
star winds \citep{ro2002}.  Perhaps the purported wind shocks could be
seeded by pulsations at the base of the wind.  Seeding the instability
at the base has been shown to lead to stronger shocks and more X-ray
emission \citep{Feldmeier1997}, though the shocks still propagate
relatively rapidly away from the star.

Another problem with this scenario is the very large X-ray production
efficiency that is implied.  Again, this general problem -- that early
B stars produce a lot of X-rays given their weak winds -- has been
known for quite some time.  Quantifying it here, using the outer wind
shock scenario presented above, and the mass-loss rate of $\Mdot
\approx 10^{-9}$ \Msunyr\/ implied by this analysis, an appreciable
fraction of the wind material is heated to X-ray emitting temperatures
at any given time.  And since the wind must be nearly completely
decelerated when it passes through the shock front, a similar fraction
of the available wind kinetic energy would go into heating the wind to
$T \ga 10^6$ K. The self-excited LDI typically is much less efficient
than this \citep{Feldmeier1997}.

If this scenario of a nearly completely shocked-heated outer wind with
a low velocity is not likely, then what are the alternatives?  As
already stated, a more standard model of embedded wind shocks formed
by the LDI and outflowing with the wind cannot explain the very modest
X-ray emission line widths nor the relatively large emission measures
(or shock heating efficiency).  However, a standard coronal-type
scenario, as applied successfully to cool stars, cannot explain the
data either.  Coronal sources do not have emission lines that can be
resolved in the \chandra\/ \hetgs.  And the \ftoi\/ ratios would be
lower than observed if the hot plasma were magnetically confined near
the surface of the star.  And of course, there is no reasonable
expectation of a strong dynamo in OB stars. The relative softness and
lack of variability would also seem to argue against an origin for the
X-rays in a traditional corona at the base of the wind.

Perhaps magnetic fields are involved, but just not in the context of a
dynamo and corona. Magnetic channeling of hot-star winds has been
shown to efficiently produce X-rays and that X-ray emission has only
modestly broad emission lines \citep{Gagne2005}.  There is, however,
no indication that \bcru\/ has a magnetic field.  However, a field of
only $B \approx 5$ G would be sufficiently strong to channel the
low-density wind of \bcru. The degree of magnetic confinement and
channeling can be described by the parameter

\[
\eta_{\ast} \equiv \frac{B^2R_{\ast}^2}{\Mdot\vinf}, 
\]

\noindent
where values of $\eta_{\ast} > 1$ imply strong magnetic confinement
and $B$ is the equatorial field strength for an assumed dipole field
\citep{uo2002}.  We note that a stronger field has already been
detected in another \bcep\/ star, $\beta$ Cep itself
\citep{Donati2002}, though attempts to detect a field on \bcru\/ have
not yet been successful \citep{Hubrig2006}.  The upper limits on the
field strength are roughly an order of magnitude higher than what
would be needed for significant confinement and channeling, though. We
also note that the radius of the last closed loop scales as
$\eta_{\ast}^{0.5}$ \citep{uo2002}, so a field of 50 G -- still
undetectable -- would lead to a magnetosphere with closed field lines
out to 3 \Rstar, consistent with the value found for \Rfir in
\S4.1.3.

In a more speculative scenario, it is possible that the material
giving rise to the X-ray emission is near the base of the wind, and is
falling back toward the star -- a failed wind.  This scenario has been
invoked to explain the X-ray emission from $\tau$ Sco (B0.2 V)
\citep{Howk2000}, which subsequently had a complex surface magnetic
field detected on it \citep{Donati2006}. If material gets accelerated
off the surface by radiation pressure, but after moving a distance off
the surface finds that it can no longer be accelerated -- perhaps
because of the formation of optically thick clumps - those clumps may
stall or even fall back toward the star and interact with the ambient
wind, leading to shock heating of wind material and material on the
surfaces of the clumps. The problems with this scenario include the ad
hoc nature of the clump formation and wind fall-back (though this
behavior is in fact seen in conjunction with magnetic channeling
\citep{uo2002,Gagne2005}) and the \ftoi\/ ratios which indicate that
the hot material is several stellar radii above the surface.
\chandra\/ observations of \tsco\/ do show \ftoi\/ ratios consistent
with hot plasma significantly above the photosphere and they also show
only modestly broadened X-ray emission lines, similar to what we see
in \bcru, although the spectrum of \tsco\/ is significantly harder
\citep{Cohen2003}.

Finally, perhaps the physical characteristics of a wind with a slow,
hot outer component that we described in our initial scenario is
correct -- that is what the \ftoi\/ ratios and the line widths are
telling us -- but the heating mechanism is not LDI-related wind
shocks.  Instead a more complete plasma treatment may be required to
describe at least the outer wind, where the densities are below $10^8$
cm$^{-3}$. In such low density winds, the radiatively driven ions can
decouple from the protons and rapidly accelerate, leading to
frictional heating \citep{sp1992,go1994,kk2001,op2002}.  Heating to
X-ray emitting temperatures is predicted to occur only at about
spectral type B5 \citep{kk2001}, but perhaps if the wind mass-loss
rates of B stars are lower than anticipated significant frictional
heating could occur in early B stars, like \bcru.

We point out that the energy requirements for heating the low density
($n < 10^8$ cm$^{-3}$) outer wind of \bcru\/ are not severe at all.
The heating mechanism might even involve wave propagation if there is
even a weak magnetic field on this star. And once significantly
heated, such low density circumstellar matter will tend to remain hot,
due to the inefficiency of radiative cooling \citep{Owocki2004}.  In
this sense, the hot plasma in the wind of \bcru\/ could be considered
a corona, and though likely initiated by radiation pressure as
described in CAK theory, the low-density, slow, far wind of \bcru\/
might even be driven by gas pressure gradients.

None of these scenarios of X-ray production on \bcru\/ are both
physically natural and also in agreement with all the available data.
Although, if a magnetic field were to be detected on \bcru, then the
magnetically channeled wind shock interpretation would be quite
reasonable.  However, we stress that the main characteristics that
constrain any models that may be put forward are quite secure: the
plasma temperature is several million degrees; the emission is not
highly variable, except for the periodic hard X-ray variability; the
X-ray emission lines are broadened but only modestly so; and the X-ray
production efficiency (if the X-ray emission is related to the star's
wind) is rather high.  Finally, we note that the periodic signal in
the hard X-rays and its phasing with the radial velocity variability
might be a clue to the X-ray production mechanism. 

Put in the context of the empirical X-ray spectral properties of OB
stars, the case of \bcru\/ may be extreme but it seems to lie at one
end of a continuum of line-profile behavior in hot stars with
radiation-driven stellar winds.  The late-type O stars $\delta$ Ori
\citep{Miller2002} and $\sigma$ Ori A \citep{Skinner2004,Skinner2008}
have lines that are broad, but not as broad as would be expected in
the standard wind-shock scenario given their known wind properties.
Perhaps there is a gradual transition from stars with X-ray emission
lines consistent with their fast stellar winds, such as \zpup\/
\citep{Kahn2001,Cassinelli2001,kco2003} and \zori\/ \citep{Cohen2006},
through some late O stars with X-ray lines of intermediate width, to
early B stars like \bcru\/ for which \chandra\/ can barely resolve
their lines.

The newly discovered companion to \bcru\/ is most likely a low-mass
pre-main-sequence star, similar to the post T Tauri stars found in the
\rosat\/ data \citep{pf1996,Alcala2002}. It has a hard, thermal
spectrum as PMS stars do, and shows significant variability, again, as
PMS stars do \citep{Favata2005}. The age of \bcru\/ is less than the
pre-main-sequence lifetimes of most low-mass stars.
\citet{Lindroos1986} identified a few dozen B stars with likely PMS
companions.  If this newly discovered companion indeed is a PMS star,
then \bcru\/ would be a Lindroos binary. As such, it would be useful
for testing evolutionary models if an optical spectrum of the
companion can be obtained.  It would be especially helpful if enough
information about the mid-B spectroscopic companion (which is supposed
to still be on the main sequence \citep{Aerts1998}) can also be
obtained.  There are systematic age differences as a function of
spectral type in the LCC which seem to reflect problems with models
\citep{PreibischMamajek2007}, and the analysis of the three coeval
stars in the \bcru\/ system could provide important constraints on the
evolutionary models.

Given the companion's X-ray luminosity of nearly $10^{30}$ ergs
s$^{-1}$ cm$^{-2}$ at the assumed distance of 108 pc, even a G star
somewhat above the main sequence would have a $f_X/f_V$ ratio
consistent with those seen in PMS stars.  The nearby K and M post-T
Tauri stars detected in the \rosat\/ pointing are typically
$m_{\mathrm v} \approx 13$ and have $f_X/f_V \approx 10^{-2}$
\citep{pf1996}. If the companion has a similar X-ray-to-V-magnitude
flux ratio, then it is a $12^{th}$ magnitude PMS K star.  If the
magnitude difference between \bcru\/ and this newly discovered
companion is roughly 11, then at a separation of 4\arcsec\/ and with
the primary being so bright, detecting it with a ground-based optical
or IR telescope will be a challenge, though such data would obviously
be very useful for confirming its PMS status and characterizing its
properties.  Any accretion disk it might have would be significantly
irradiated by UV light from the primary. Finally, given a projected
separation of 430 AU, the companion should have an orbital period
(around \bcru\/ and its spectroscopic companion) of at least 1000
years.

\section*{Acknowledgments}

Support for this work was provided by the National Aeronautics and
Space Administration through \chandra\/ Award Numbers GO2-3030A and
AR7-8002X to Swarthmore College, issued by the \chandra\/ X-ray
Observatory Center, which is operated by the Smithsonian Astrophysical
Observatory for and on behalf of the National Aeronautics and Space
Administration under contract NAS8-03060.  MAK was supported by a
Eugene M. Lang Summer Research Fellowship from the Provost's Office at
Swarthmore College.  The authors thank Stan Owocki and Ken Gayley for
fruitful discussions, especially about the physics of low density
winds and the role of pulsations, and Maurice Leutenegger for sharing
his insights about analyzing grating spectra and for the use of his
XSPEC models which we employed for modeling the He-like \ftoi\/ ratios
and the wind-broadened lines. We thank Eric Mamajek for discussions
about the LCC.  We thank Conny Aerts and Jan Cuypers for discussions
about \bcep\/ variables and for sharing unpublished data with us. We
thank Dave Huenemoerder and John Houck for their help modifying the
ISIS code to account for altered \ftoi\/ ratios. We also thank Mark
Janoff and Mary Hui for some preliminary work on the data analysis.
And we thank the referee, Ian Howarth, for his careful reading and
many useful comments and suggestions.  This research has made use of
the SIMBAD database and the Vizier catalog access tool, operated at
CDS, Strasbourg, France, and it has also made use of NASA's
Astrophysics Data System Bibliographic Services.






\newpage

\end{document}